\documentclass[lettersize,journal]{IEEEtran}
\usepackage{amsmath,amsfonts}
\usepackage{array}
\usepackage{textcomp}
\usepackage{stfloats}
\usepackage{url}
\usepackage{verbatim}
\usepackage{comment}
\usepackage{cite}
\usepackage{lipsum}
\usepackage[table]{xcolor}
\usepackage{colortbl}

\usepackage{bm}  
\usepackage{float} % For [H] option
\usepackage{placeins} % For FloatBarrier
\usepackage{amsmath,amssymb,amsfonts}
\usepackage{graphicx}
\usepackage{textcomp}
\usepackage{xcolor}
\usepackage{caption}
\usepackage{cite}
\usepackage{multirow}
\usepackage{url}
\usepackage{amsmath,amssymb,amsfonts}
\usepackage{graphicx}
\usepackage{textcomp}
\usepackage{caption}
\usepackage{multicol}
\usepackage{xcolor}
\usepackage{subcaption}
\usepackage{array}
\usepackage{float}

\usepackage{algorithm}
\usepackage{algorithmic}
\usepackage{graphicx}
\usepackage{textcomp}
\usepackage{scalerel}
\usepackage{cite}
\usepackage{amsmath,amssymb,amsfonts}
\usepackage{algorithmic}
\usepackage{graphicx}
\usepackage{textcomp}
\usepackage{caption}
\usepackage{multicol}
\usepackage{xcolor}
\usepackage{esdiff}
\usepackage{hyperref}
\usepackage{cleveref}
\usepackage{etoolbox}
\usepackage{cases}
\usepackage{mathtools}
\usepackage{comment}
\usepackage{amssymb}     % checkmarks
\usepackage{pifont}      % ding symbols for ✓ and ✗
\usepackage{booktabs}    % professional table rules
\newcommand{\cmark}{\ding{51}} % ✓
\definecolor{headergray}{RGB}{235,235,235}
\definecolor{smallscale}{RGB}{225,240,255}   % light blue
\definecolor{mediumscale}{RGB}{235,255,235}  % light green
\definecolor{largescale}{RGB}{255,235,235}   % light red

\renewcommand{\baselinestretch}{1}

\usepackage[none]{hyphenat} % Reduce hyphenation
\usepackage{microtype}      % Enhance justification
\begin{document}
	
\title{All-Pass \textcolor{black}{Function} OPF: A Solver-Friendly, Physics-Preserving Approximation of AC OPF}
	
\author{Milad Hasanzadeh,~\textit{Graduate Student Member, IEEE},
        Amin Kargarian,~\textit{Senior Member, IEEE},
        and Javad Lavaei,~\textit{Fellow, IEEE}
\thanks{This work was supported by the National Science Foundation under 
Grant ECCS-1944752 and Grant ECCS-2312086.}
\thanks{M.~Hasanzadeh and A.~Kargarian are with the Department of Electrical and Computer Engineering,
Louisiana State University, Baton Rouge, LA 70803 USA 
(e-mail: mhasa42@lsu.edu; kargarian@lsu.edu).}
\thanks{J.~Lavaei is with the Department of Industrial Engineering and Operations Research,
University of California, Berkeley, CA 94720 USA
(e-mail: lavaei@berkeley.edu).}
}

\maketitle

\begin{abstract}
This paper presents a \textcolor{black}{rational} approximation of the AC optimal power flow (AC OPF) problem based on an all-pass approximation of the exponential power flow kernel. The classical AC OPF relies on trigonometric coupling between bus voltage phasors, which yields a nonconvex program with oscillatory derivatives that can slow, or in some cases destabilize, interior-point methods. We replace the trigonometric terms with an all-pass \textcolor{black}{function} (APF) approximation whose real and imaginary components act as smooth surrogates for the cosine and sine functions, and we introduce a pre-rotation to shift the argument of the approximation toward its most accurate region, ensuring that the reformulated power flow model preserves physical loss behavior, maintains the symmetry of the classical kernels, and improves the conditioning of the Jacobian and Hessian matrices. The proposed APF OPF formulation remains nonconvex, as in the classical model, but it eliminates trigonometric evaluations and empirically produces larger and more stable Newton steps under standard interior-point solvers. Numerical results on more than 25 IEEE and PGLib test systems ranging from 9 to 10{,}000 buses demonstrate that the APF OPF model achieves solutions with accuracy comparable to that of the classical formulation while reducing solver times, indicating a more solver-friendly nonconvex representation of AC OPF. All code, functions, verification scripts, and generated results are publicly available on \href{https://github.com/LSU-RAISE-LAB/APF-OPF}{GitHub}, along with a README describing how to run and reproduce the experiments.
\end{abstract}


\section*{Nomenclature}
\label{sec:nomenclature}
\addcontentsline{toc}{section}{Nomenclature}

\setlength{\IEEElabelindent}{2em}
\setlength{\labelsep}{1.2em}

\begin{IEEEdescription}[
  \IEEEusemathlabelsep\IEEEsetlabelwidth{$\mathrm{APF}_2(\delta_{ij})$}
]

% ============================================================
% Sets
% ============================================================
\item[\emph{Sets}]
\item[$\mathcal{N}$] Set of buses.
\item[$\mathcal{G}$] Set of generators.
\item[$\mathcal{G}_i$] Set of generators located at bus $i$.
\item[$\mathcal{L}$] Set of transmission lines.

% ============================================================
% Parameters
% ============================================================
\item[\emph{Parameters}]
\item[$\mathbf{j}$] Imaginary unit, $\mathbf{j} := \sqrt{-1}$.
\item[$p^d_i, q^d_i$] Active/reactive load at bus $i$.
\item[$\underline{p}^g_i, \overline{p}^g_i$] Lower/upper active power generation limits.
\item[$\underline{q}^g_i, \overline{q}^g_i$] Lower/upper reactive power generation limits.
\item[$\underline{V}_i, \overline{V}_i$] Lower/upper voltage magnitude limits.
\item[$Y_{ij}$] Line admittance between buses $i$ and $j$.
\item[$|Y_{ij}|$] Magnitude of admittance entry $Y_{ij}$.
\item[$\phi_{ij}$] Phase angle of admittance entry $Y_{ij}$.
\item[$\underline{\Delta\theta}_{ij}, \overline{\Delta\theta}_{ij}$] Angle-difference limits.
\item[$\overline{S}_{ij} $] Thermal rating of branch $(i,j)$.
\item[$f(p_i^g)$] Generation cost function.
\item[$\theta_{\mathrm{ref}}$] Reference (slack) bus angle.
\item[$a$] All-pass design parameter.
\item[$\theta_i^{dc}$] DC power flow voltage angle at bus $i$.
\item[$\delta_{ij}^{dc}$] DC angle difference on branch $(i,j)$.
\item[$\mu$] Barrier parameter for interior-point method.
\item[$\alpha$] Step size in Newton/IPM updates.

% ============================================================
% Variables
% ============================================================
\item[\emph{Variables}]
\item[$p^g_i, q^g_i$] Active/reactive generation at bus $i$.
\item[$V_i$] Voltage magnitude at bus $i$.
\item[$\theta_i$] Voltage angle at bus $i$.
\item[$\delta_{ij}$] AC angle difference $\theta_i - \theta_j - \phi_{ij}$.
\item[$P_{ij}, Q_{ij}$] Power flow from bus $i$ to bus $j$.
\item[$P_{ji}, Q_{ji}$] Power flow from bus $j$ to bus $i$.
\item[$r(\delta)$] Complex all-pass kernel.
\item[$r_{\cos}(\delta)$] Real part of $r(\delta)$; surrogate for $\cos(\delta)$.
\item[$r_{\sin}(\delta)$] Imaginary part of $r(\delta)$; surrogate for $\sin(\delta)$.
\item[$s$] Slack variables for inequality constraints.
\item[$\lambda$] Lagrange multipliers for equality constraints.
\item[$\nu$] Lagrange multipliers for inequality constraints.

\end{IEEEdescription}

\section{Introduction}
\label{sec:introduction}

\IEEEPARstart{T}{he} AC optimal power flow (AC OPF) is a fundamental optimization problem in power system analysis, operation, and planning \cite{frank2016introduction}. It aims to minimize total generation cost while satisfying the nonlinear AC power flow equations and operational constraints such as generator limits, voltage bounds, branch thermal limits, and phase-angle restrictions \cite{low2014convex}. By explicitly modeling network physics through complex voltage phasors, AC OPF captures real and reactive power flows, losses, and voltage interactions via trigonometric coupling between voltage magnitudes and angles \cite{molzahn2019survey,hasanzadeh2025admm2}. However, this nonlinear formulation leads to a highly nonconvex feasible set, making AC OPF computationally challenging, particularly for large-scale and heavily loaded transmission systems \cite{bienstock2019strong,low2014convex, molzahn2019survey,hasanzadeh2025admm}.

In general, research on AC OPF addresses its computational challenges through two main directions: relaxations and approximations \cite{molzahn2019survey}. Relaxation-based methods mitigate nonconvexity by reformulating or convexifying the original problem, often yielding optimization models that are easier to solve and may provide global or near-global optimality guarantees \cite{madani2014convex,bose2014equivalent,coffrin2015strengthening,jabr2006radial}. In contrast, approximation-based methods preserve the original OPF structure but simplify the AC power flow equations using linearized or reduced-complexity models, reducing computational effort at the expense of accuracy \cite{buason2024adaptive,goodwin2025power,stott2009dc}. Although both approaches address nonlinear network physics, relaxations emphasize mathematical tractability and optimality bounds, whereas approximations prioritize physical interpretability and scalability for large-scale applications \cite{molzahn2019survey}.

Relaxation-based approaches to AC OPF have been widely studied as a way to address the inherent nonconvexity of the problem. A major line of work focuses on convex relaxations, where the nonlinear AC power flow equations are reformulated into convex optimization problems, such as semidefinite programming \cite{lavaei2011zero}, second-order cone programming \cite{farivar2013branch,low2014convex}, or quadratic convex \cite{coffrin2015strengthening} relaxations. These methods replace the original nonconvex feasible set with a convex superset, allowing global or near-global optimal solutions to be computed under certain conditions \cite{lavaei2011zero,low2014convex}. When the relaxation is exact, the solution of the relaxed problem coincides with that of the original AC OPF, a phenomenon that has been demonstrated to occur for specific network topologies and operating regimes.

Approximation-based approaches to AC OPF reduce computational complexity by simplifying the nonlinear AC power flow equations. A classic example is the DC OPF model, which neglects reactive power, voltage magnitude variations, and losses, resulting in a linear and computationally efficient representation of network behavior \cite{stott2009dc}. To improve accuracy beyond DC OPF, various linearized and nonlinear approximations have been proposed, including Taylor-based expansions, fixed-point methods, and other reduced-complexity representations of trigonometric power flow relations \cite{molzahn2019survey,hasanzadeh2026benchmarklibrarydistributedpower}. More recent work has explored adaptive and data-driven approximation techniques that better capture nonlinear AC behavior across different operating conditions \cite{buason2024adaptive,goodwin2025power}. In particular, \textcolor{black}{rational} approximations based on Pad\'e functions have been shown to achieve higher accuracy than DC OPF, and have been used to derive convex OPF formulations with improved fidelity \cite{buason2024adaptive}. Despite these advances, the accuracy of approximation-based methods generally depends on the operating point, and preserving consistency with physical power flow properties remains a key challenge \cite{molzahn2019survey}.

From a computational perspective, the relevance of approximation choices becomes evident when considering how AC OPF problems are solved in practice. AC OPF is most commonly addressed using nonlinear programming solvers, among which interior-point methods (IPMs) have emerged as the dominant approach. IPMs solve constrained optimization problems by iteratively minimizing a sequence of barrier-augmented subproblems, where inequality constraints are incorporated into the objective through logarithmic barrier terms, and the resulting optimality conditions are solved using Newton-type steps \cite{wachter2006implementation,hasanzadeh2024distributed,nocedal2006numerical}. Due to their favorable scaling behavior and robustness for large nonlinear systems, IPMs are widely used in practical power system tools. In fact, the MATPOWER package adopts an IPM-based solver, MIPS, as its default and most reliable solver for AC OPF \cite{zimmerman2010matpower,zimmerman2016matpower}.

Despite their effectiveness, the performance of IPMs is influenced by the smoothness and conditioning of the underlying model equations. In AC OPF, the trigonometric coupling between voltage magnitudes and phase angles produces highly nonlinear and oscillatory derivatives, which can result in ill-conditioned Jacobian and Hessian matrices, small Newton step sizes, and slow or unstable convergence, especially in large-scale or heavily loaded networks \cite{molzahn2019survey}. These challenges motivate the development of alternative AC OPF formulations that retain physical accuracy while offering smoother and more solver-friendly derivative structures. In particular, approximation models that align naturally with the numerical characteristics of IPMs provide a promising direction.

These observations motivate the use of \textcolor{black}{rational} approximations not only as accuracy-enhancing tools, but also as a means of re-modeling the nonconvex AC power flow equations in a form that is more compatible with Newton-based optimization algorithms. By reshaping the nonlinear trigonometric relationships into \textcolor{black}{rational} functions, rational models can improve derivative smoothness and conditioning, which is critical for robust IPM performance. Among \textcolor{black}{rational} approximations, Pad\'e approximations of complex exponential functions do not preserve unit magnitude and may introduce poles or near-singular behavior. When embedded within power flow equations, this will lead to artificial gain or attenuation, distorting the physical representation of power transfer and losses and potentially degrading numerical stability during optimization. These limitations highlight the need for \textcolor{black}{rational} approximations that preserve both numerical robustness and fundamental physical properties of AC power flow models.

Motivated by the numerical challenges of solving AC OPF with interior-point methods and the limitations of classical polynomial approximations, this paper presents an all-pass \textcolor{black}{function} (APF) approximation of the AC power flow equations for OPF applications, termed APF OPF. The goal is not to eliminate the inherent nonconvexity of AC OPF, but to obtain a milder, more solver-friendly nonconvex representation by replacing the trigonometric sine and cosine terms with smooth APF functions. These APF functions preserve unit magnitude and the symmetry of the original power flow model, allowing the physical interpretation of power transfer and losses to be maintained. To improve approximation accuracy in practical operating regions, a pre-rotation of voltage angle differences based on a DC power flow solution is introduced, which centers the \textcolor{black}{rational} approximation around typical angle values encountered during optimization. As a result, the APF OPF formulation retains the structure of the AC OPF model while exhibiting smoother and better-conditioned Jacobian and Hessian matrices that are suitable for interior-point solvers. Numerical experiments on more than 25 IEEE and PGLib test systems (9 to 10{,}000 buses) show that APF-OPF matches the accuracy of classical AC-OPF while reducing solver time, offering a more tractable nonconvex alternative in practice.

\section{AC OPF: Modeling and Interior-Point Methods}\label{sec:acopf}
The standard AC OPF formulation is given in \eqref{eq:acopf_exp}, where the decision variables are \(V\), \(\theta\), \(p^g\), and \(q^g\).

\subsection{AC OPF Model}
\begin{subequations}\label{eq:acopf_exp}
\begin{align}
\min_{\{V,\theta,p^g,q^g\}} \;
& \sum_{i\in\mathcal{G}} f(p_i^g)
\label{eq:acopf_exp_obj}
\\
\text{s.t.} \quad
& \sum_{k\in\mathcal{G}_i} p_k^g - p_i^d
  = \!\! \sum_{j\in\mathcal{N}} \! V_iV_j|Y_{ij}|
    \cos(\delta_{ij}),
\label{eq:acopf_exp_Pbal}
\\
& \sum_{k\in\mathcal{G}_i} q_k^g - q_i^d
  = \!\! \sum_{j\in\mathcal{N}} \! V_iV_j|Y_{ij}|
    \sin(\delta_{ij}),
\label{eq:acopf_exp_Qbal}
\\
& \underline{p}_i^g \le p_i^g \le \overline{p}_i^g,\quad
  \underline{q}_i^g \le q_i^g \le \overline{q}_i^g,
\label{eq:acopf_exp_gen}
\\
& \underline{V}_i \le V_i \le \overline{V}_i,
\label{eq:acopf_exp_V}
\\
& -\pi \le \theta_i \le \pi,\qquad
  \theta_{\text{ref}}=0,
\label{eq:acopf_exp_theta}
\\
& \underline{\Delta\theta}_{ij}
  \le \theta_i - \theta_j
  \le \overline{\Delta\theta}_{ij},
\quad \forall (i,j)\!\in\!\mathcal{L},
\label{eq:acopf_exp_ang}
\\
& P_{ij}^2(V,\theta)+Q_{ij}^2(V,\theta)
  \le \overline{S}_{ij}^2 ,
\label{eq:acopf_exp_thermal_f}
\\
& P_{ji}^2(V,\theta)+Q_{ji}^2(V,\theta)
  \le \overline{S}_{ij}^2 .
\label{eq:acopf_exp_thermal_t}
\end{align}
\end{subequations}
\noindent
where $\delta_{ij} := \theta_i - \theta_j - \phi_{ij}$. The optimization problem \eqref{eq:acopf_exp} minimizes the total generation cost \eqref{eq:acopf_exp_obj}. The active and reactive nodal balances \eqref{eq:acopf_exp_Pbal}–\eqref{eq:acopf_exp_Qbal} enforce Kirchhoff’s laws using the trigonometric power flow kernel derived from the Y-bus. Generator limits are imposed by \eqref{eq:acopf_exp_gen}, voltage bounds by \eqref{eq:acopf_exp_V}, and angular feasibility by \eqref{eq:acopf_exp_theta}–\eqref{eq:acopf_exp_ang}. \eqref{eq:acopf_exp_thermal_f}–\eqref{eq:acopf_exp_thermal_t} enforce thermal limits on each transmission line.

The AC OPF problem is nonlinear because the power flow relations involve products of voltage magnitudes $V_iV_j$ and nonlinear trigonometric functions $\cos(\delta_{ij})$ and $\sin(\delta_{ij})$. The problem is also nonconvex. The cosine and sine terms create feasible regions that are not convex sets. This non-convexity is a fundamental challenge of AC OPF and is the reason why advanced solvers are used.

The AC power flow equations are symmetric with respect to phase-angle shifts: adding a constant to all $\theta_i$ does not change any $\delta_{ij}$ or any power flows. The underlying kernel also has unit magnitude, meaning
\[
|\cos(\delta_{ij}) + \mathbf{j}\sin(\delta_{ij})| = 1.
\]
Thus, the exponential kernel used in AC power flow does not introduce artificial losses—the model preserves the physical property that only network resistances contribute to real losses, preventing accumulation of spurious losses that would distort feasibility or optimality.

\subsection{Interior-Point Solution}

Let $x := \{\, V_i,\;\theta_i,\; p_i^g,\; q_i^g \,\}_{i\in\mathcal{N}}$ collect all decision variables. The power-balance constraints \eqref{eq:acopf_exp_Pbal}–\eqref{eq:acopf_exp_Qbal} form the nonlinear equality constraints, and generator, voltage, angle-difference, and thermal limits in \eqref{eq:acopf_exp_gen}–\eqref{eq:acopf_exp_thermal_t} form the inequality constraints. Therefore, AC OPF can be written compactly as
\begin{subequations}\label{eq:nlp_acopf}
\begin{align}
\min_{x} \;& f(x), \\
\text{s.t.} \;& h(x)=0, \\
            & g(x)\le 0,
\end{align}
\end{subequations}
where $h(x)$ enforces active and reactive nodal balances, and $g(x)$ aggregates all remaining operational limits. \textcolor{black}{IPM introduces positive slack variables $s$ to
write the inequality constraints as $g(x)+s=0$. For barrier
parameter $\mu>0$, the barrier-augmented objective becomes}
\begin{align}
{\color{black}
\Phi(x,s,\mu)=f(x)-\mu\sum_i\ln(s_i),}
\label{eq:acopf_barrier}
\end{align}
\textcolor{black}{and the corresponding subproblem is}
\begin{align}
{\color{black}
\begin{aligned}
\min_{x,s}\;&\Phi(x,s,\mu)\\
\text{s.t.}\;&h(x)=0,\quad g(x)+s=0,\quad s>0.
\end{aligned}}
\label{eq:acopf_barrier_sub}
\end{align}
\textcolor{black}{The barrier Lagrangian is}
\begin{align}
{\color{black}
\begin{aligned}
L(x,s,\lambda,\nu,\mu)
={}&f(x)-\mu\sum_i\ln(s_i)\\
&+\lambda^T h(x)+\nu^T(g(x)+s),
\end{aligned}}
\label{eq:acopf_lagr}
\end{align}
\textcolor{black}{where $\lambda$ and $\nu$ are the multipliers
associated with $h(x)=0$ and $g(x)+s=0$, respectively.
The first-order optimality conditions are}
\begin{subequations}\label{eq:acopf_kkt}
{\color{black}
\begin{align}
\nabla_x L(x,s,\lambda,\nu,\mu)&=0,\\
h(x)&=0,\\
g(x)+s&=0,\\
\nu_i s_i&=\mu.
\end{align}}
\end{subequations}
\textcolor{black}{Here, $s>0$ and $\nu>0$. The last condition
follows from stationarity with respect to $s_i$,
$\nu_i-\mu/s_i=0$. These relations encode primal feasibility,
dual feasibility, and perturbed complementarity.}

Because the active and reactive power-balance constraints involve trigonometric expressions, IPM needs analytic derivatives such as $\frac{\partial P_i}{\partial V_k},$ and $\frac{\partial P_i}{\partial \theta_k}, $ obtained by differentiating the kernel
\[
P_i = \sum_j V_iV_j|Y_{ij}| \cos(\delta_{ij}).
\]
These derivatives populate the constraint Jacobians
\begin{align}
J_h(x) = \nabla h(x)^T, 
\qquad
J_g(x) = \nabla g(x)^T,
\label{eq:jac_mid}
\end{align}
and the Hessian of the barrier Lagrangian
\begin{align}
{\color{black}
H(x,\lambda,\nu)
= \nabla_{xx}^{2}L(x,s,\lambda,\nu,\mu).}
\end{align}

At each IPM iteration, Newton’s method linearizes the optimality system \eqref{eq:acopf_kkt}. \textcolor{black}{Let $r(x,\lambda,\nu,s,\mu)$ denote the concatenated stationarity, primal feasibility, and
complementarity residuals.} The Newton direction $(\Delta x,\Delta\lambda,\Delta\nu,\Delta s)$ solves
\begin{align}
K
\begin{bmatrix}
\Delta x \\[0.5mm]
\Delta\lambda \\[0.5mm]
\Delta\nu \\[0.5mm]
\Delta s
\end{bmatrix}
=
-r(x,\lambda,\nu,s,\mu),
\label{eq:acopf_newton}
\end{align}
where the KKT matrix is
\begin{align}
K =
\begin{bmatrix}
H & J_h^T & J_g^T & 0 \\
J_h & 0 & 0 & 0 \\
J_g & 0 & 0 & I \\
0 & 0 & S & N
\end{bmatrix},
\label{eq:acopf_kkt_matrix}
\end{align}
with $S=\operatorname{diag}(s)$ and $N=\operatorname{diag}(\nu)$. The structure of $K$ reflects the entire AC OPF physics: the Jacobian captures the sensitivity of nodal balances and branch flows to voltage magnitudes and angles, while the Hessian captures curvature introduced by trigonometric coupling.

Once the Newton direction is computed, a step size is chosen to maintain $s>0$ and $\nu>0$, and primal and dual variables are updated via
\[
x^{(k+1)} = x^{(k)} + \alpha \Delta x, \qquad
\lambda^{(k+1)} = \lambda^{(k)} + \alpha \Delta\lambda,
\]
\[
\nu^{(k+1)} = \nu^{(k)} + \alpha \Delta\nu, \qquad
s^{(k+1)} = s^{(k)} + \alpha \Delta s.
\]
As $\mu$ decreases, the iterates converge to a Karush–Kuhn–Tucker (KKT) point of AC OPF \cite{bertsekas1997nonlinear}. Interior-point performance, therefore, depends critically on the conditioning and smoothness of the Jacobians and Hessians derived from the AC power flow equations. Poorly conditioned derivatives—especially from the trigonometric kernel—can degrade Newton steps, slow convergence, or cause numerical instability, motivating the development of approximations with a more solver-friendly structure.

\section{Proposed APF OPF and Its Properties}
\label{sec:allpass}
We present the APF approximation of the trigonometric kernel of the power flow equations and discuss its properties in terms of (i) preserving the physical laws governing the electrical system and (ii) improving computational efficiency.

\subsection{All-Pass Kernel Approximation}

The classical AC power flow model is built on the complex kernel
\begin{align}
e^{j\delta} = \cos(\delta) + \mathbf{j} \sin(\delta),
\label{eq:exp_kernel}
\end{align}
whose real and imaginary parts represent active and reactive coupling between bus voltage phasors. The magnitude of this kernel is exactly one, which ensures that only the network resistances contribute to real power losses. This unit-magnitude property is important because it prevents the artificial amplification or damping of voltage phasors as they propagate through the network.

To approximate the exponential kernel \eqref{eq:exp_kernel} in a rational form, inspired by the concept of all-pass filters in signal processing \cite{oppenheim1997signals}, we use a first-order all-pass function
\begin{align}
r(\delta) = \frac{1 + \mathbf{j} a \delta}{1 - \mathbf{j} a \delta},
\label{eq:allpass_def}
\end{align}
where \(a > 0\) is a design parameter controlling the shape of the approximation. The function \(r(\delta)\) is called all-pass because its magnitude is identically one for all real \(\delta\). This can be verified by computing
\begin{align}
|r(\delta)|^2
&= \frac{(1 + \mathbf{j} a \delta)(1 - \mathbf{j} a \delta)}{(1 - \mathbf{j} a \delta)(1 + \mathbf{j} a \delta)} = \frac{1 + (a\delta)^2}{1 + (a\delta)^2} = 1.
\label{eq:allpass_mag_one}
\end{align}

To relate \eqref{eq:allpass_def} to trigonometric functions, we write its real and imaginary parts explicitly. Multiplying numerator and denominator by \(1 + \mathbf{j} a \delta\) yields
\begin{align}
r(\delta)
= \frac{1 + \mathbf{j} a \delta}{1 - \mathbf{j} a \delta}
= \frac{(1 + \mathbf{j} a \delta)^2}{1 + (a\delta)^2}.
\end{align}
Expanding the numerator gives
\begin{align}
(1 + \mathbf{j} a \delta)^2
= 1 + 2 \mathbf{j} a \delta - (a\delta)^2,
\end{align}
so that
\begin{align}
r(\delta)
= \frac{1 - (a\delta)^2}{1 + (a\delta)^2}
+ \mathbf{j} \frac{2 a \delta}{1 + (a\delta)^2}.
\label{eq:allpass_re_im}
\end{align}
This suggests the following approximations:
\begin{align}
\cos(\delta)
&\approx r_{\cos}(\delta)
:= \frac{1 - (a\delta)^2}{1 + (a\delta)^2},
\label{eq:rcos_def}
\\
\sin(\delta)
&\approx r_{\sin}(\delta)
:= \frac{2 a \delta}{1 + (a\delta)^2}.
\label{eq:rsin_def}
\end{align}
By choosing \(a\) appropriately, one can match the local slope of \(\sin(\delta)\) near the expansion point and obtain a good approximation in a neighborhood around that point. At the same time, the unit-magnitude property \(|r(\delta)| = 1\) guarantees that the rational kernel preserves the loss behavior of the original exponential kernel.

\subsection{Pre-Rotation}

The first-order all-pass functions in \eqref{eq:rcos_def}–\eqref{eq:rsin_def} approximate $\sin(\delta)$ and $\cos(\delta)$ best when their argument is close to zero. However, in a real power system, the electrical angle differences
\[
\delta_{ij} = \theta_i - \theta_j - \phi_{ij}
\]
are not necessarily small. They depend on loading conditions and the network structure, so directly applying the all-pass approximation to $\delta_{ij}$ does not always give a good match.

To address this, we implement a pre-rotation step and shift the angle differences before applying the rational approximation. We compute a DC power flow and/or DC OPF once, obtain the DC angles $\theta_i^{dc}$, and use them as a reference point. We are not using the DC solution as an initialization for AC OPF. Instead, we only use it to pre-rotate the angle differences so that the argument of the approximation is centered around zero.

From the DC solution, the corresponding angle difference on each line $(i,j)$ is
\begin{align}
\delta_{ij}^{dc}
= \theta_i^{dc} - \theta_j^{dc} - \phi_{ij}.
\label{eq:delta_dc}
\end{align}
We now write the AC angle difference as
\begin{align}
\delta_{ij}
= \delta_{ij}^{dc} + \Delta_{ij},
\label{eq:delta_split}
\end{align}
where
\begin{align}
\Delta_{ij}
= (\theta_i - \theta_j) - (\theta_i^{dc} - \theta_j^{dc}).
\label{eq:Delta_def}
\end{align}
The term $\Delta_{ij}$ measures how far the current AC iterate is from the DC reference. In many
practical cases, the DC angles are already close to the final AC angles. That means that,
as the OPF solver converges, $\Delta_{ij}$ tends to become small:
\[
\Delta_{ij} \longrightarrow 0.
\]
This observation is important because the APF is most accurate near zero. Using the decomposition in \eqref{eq:delta_split}, we rewrite the trigonometric terms. If we define
\[
\gamma_1 = \delta_{ij}^{dc}, \qquad \gamma_2 = \Delta_{ij},
\]
then $\delta_{ij} = \gamma_1 + \gamma_2$. Applying the angle-addition formulas step by step gives
\begin{align}
\cos(\delta_{ij})
&= \cos(\gamma_1 + \gamma_2) \\
&= \cos(\gamma_1)\cos(\gamma_2) - \sin(\gamma_1)\sin(\gamma_2) \\
&= \cos(\delta_{ij}^{dc})\cos(\Delta_{ij})
  - \sin(\delta_{ij}^{dc})\sin(\Delta_{ij}),
\label{eq:cos_expand}
\end{align}
and similarly
\begin{align}
\sin(\delta_{ij})
&= \sin(\delta_{ij}^{dc})\cos(\Delta_{ij})
 + \cos(\delta_{ij}^{dc})\sin(\Delta_{ij}).
\label{eq:sin_expand}
\end{align}

The key idea is that only the terms $\cos(\Delta_{ij})$ and $\sin(\Delta_{ij})$ depend on the AC variables. Since $\Delta_{ij}$ stays small near the optimum, we can approximate them using the all-pass formulas introduced earlier:
\begin{align}
\cos(\Delta_{ij})
&\approx r_{\cos}(\Delta_{ij})
= \frac{1 - (a\Delta_{ij})^2}{1 + (a\Delta_{ij})^2},
\label{eq:rcosDelta}\\
\sin(\Delta_{ij})
&\approx r_{\sin}(\Delta_{ij})
= \frac{2a\Delta_{ij}}{1 + (a\Delta_{ij})^2}.
\label{eq:rsinDelta}
\end{align}

Substituting \eqref{eq:rcosDelta}–\eqref{eq:rsinDelta} into \eqref{eq:cos_expand}–\eqref{eq:sin_expand} gives the following approximations for each line:
\begin{align}
\cos(\delta_{ij})
&\approx
\cos(\delta_{ij}^{dc})\, r_{\cos}(\Delta_{ij})
-
\sin(\delta_{ij}^{dc})\, r_{\sin}(\Delta_{ij}),
\label{eq:APF1}
\\
\sin(\delta_{ij})
&\approx
\sin(\delta_{ij}^{dc})\, r_{\cos}(\Delta_{ij})
+
\cos(\delta_{ij}^{dc})\, r_{\sin}(\Delta_{ij}).
\label{eq:APF2}
\end{align}

To simplify notation, we define
\begin{align}
\hat{r}_{\cos}(\delta_{ij})
&:= \cos(\delta_{ij}^{dc})\, r_{\cos}(\Delta_{ij})
   -\sin(\delta_{ij}^{dc})\, r_{\sin}(\Delta_{ij}),
\\
\hat{r}_{\sin}(\delta_{ij})
&:= \sin(\delta_{ij}^{dc})\, r_{\cos}(\Delta_{ij})
   +\cos(\delta_{ij}^{dc})\, r_{\sin}(\Delta_{ij}),
\end{align}
so the AC power flow equations originally defined in \eqref{eq:acopf_exp_Pbal} and \eqref{eq:acopf_exp_Qbal} become APF power flow equations as
\begin{align}
P_i^g - P_i^d
  &= \sum_{j} V_iV_j|Y_{ij}|\, \hat{r}_{\cos}(\delta_{ij}), \\[2mm]
Q_i^g - Q_i^d
  &= \sum_{j} V_iV_j||Y_{ij}|\, \hat{r}_{\sin}(\delta_{ij}),
\end{align}
where the functions $\hat{r}_{\cos}$ and $\hat{r}_{\sin}$ depend only on the shifted angle difference $\Delta_{ij}$ and the DC reference $\delta_{ij}^{dc}$.   A similar substitution is made for the expressions defining $P_{ij}$ and $Q_{ij}$ in the thermal-limit constraints, so the entire APF OPF model becomes consistent. 

\subsection{APF OPF Model}
The resulting APF OPF is
% =========================================================
% APF OPF model (IEEEtran-friendly, matches AC-OPF style)
% =========================================================
\begin{subequations}\label{eq:apf_opf}
\begin{align}
\min_{\{V,\theta,p^g,q^g\}} \;
& \sum_{i\in\mathcal{G}} f(p_i^g)
\label{eq:apf_obj}
\\
\text{s.t.} \quad
& \sum_{k\in\mathcal{G}_i} p_k^g - p_i^d
  = \sum_{j\in\mathcal{N}} V_iV_j|Y_{ij}|\,
    \hat r_{\cos}(\delta_{ij}),
\label{eq:apf_Pbal}
\\
& \sum_{k\in\mathcal{G}_i} q_k^g - q_i^d
  = \sum_{j\in\mathcal{N}} V_iV_j|Y_{ij}|\,
    \hat r_{\sin}(\delta_{ij}),
\label{eq:apf_Qbal}
\\
& \underline{p}_i^g \le p_i^g \le \overline{p}_i^g,\quad
  \underline{q}_i^g \le q_i^g \le \overline{q}_i^g,
\label{eq:apf_gen}
\\
& \underline{V}_i \le V_i \le \overline{V}_i,
\label{eq:apf_V}
\\
& -\pi \le \theta_i \le \pi,\qquad
  \theta_{\text{ref}}=0,
\label{eq:apf_theta}
\\
& \underline{\Delta\theta}_{ij}
  \le \theta_i - \theta_j
  \le \overline{\Delta\theta}_{ij},
\quad \forall (i,j)\in\mathcal{L},
\label{eq:apf_ang}
\\
& P_{ij}^2(V,\theta;\mathrm{APF})+Q_{ij}^2(V,\theta;\mathrm{APF})
  \le \overline{S}_{ij}^2 ,
\label{eq:apf_thermal_f}
\\
& P_{ji}^2(V,\theta;\mathrm{APF})+Q_{ji}^2(V,\theta;\mathrm{APF})
  \le \overline{S}_{ij}^2 .
\label{eq:apf_thermal_t}
\end{align}
\end{subequations}

\noindent
In \eqref{eq:apf_thermal_f}--\eqref{eq:apf_thermal_t}, the branch-flow expressions $P_{ij}(\cdot;\mathrm{APF})$ and $Q_{ij}(\cdot;\mathrm{APF})$ are computed using the same network model as in the AC OPF, except that every occurrence of $\cos(\delta_{ij})$ and $\sin(\delta_{ij})$ in the line-flow formulas is replaced by $\hat r_{\cos}(\delta_{ij})$ and $\hat r_{\sin}(\delta_{ij})$, respectively, ensuring consistency between nodal balances and thermal-limit constraints.

\subsection{Properties of APF OPF Formulation}
We analyze the main structural and numerical properties of the APF OPF formulation and explain why these properties lead to improved numerical behavior without altering the underlying physics of the power flow model.

\subsubsection{Unit-Magnitude Property and Preservation of Power Losses}

One of the most important properties of the all-pass function is that its magnitude is identically equal to one for all real angle differences:
\[
|r(\delta)| = 1 \quad \text{for every real } \delta.
\]
As a result, the APF approximation does not introduce any artificial attenuation or amplification of power flow terms. In particular, real power losses continue to arise solely from the physical resistances of transmission lines, exactly as in the classical AC OPF formulation. This ensures that the approximation does not distort the loss model or alter the physical interpretation of resistive dissipation.

\subsubsection{Symmetry Preservation of Trigonometric Structure}

Another key property of the all-pass approximation is that its real and imaginary components,
\[
r_{\cos}(\delta) = \frac{1 - (a\delta)^2}{1 + (a\delta)^2},
\qquad
r_{\sin}(\delta) = \frac{2a\delta}{1 + (a\delta)^2},
\]
preserve the symmetry of the trigonometric functions they replace. Evaluating these expressions at \(-\delta\) yields
\[
r_{\cos}(-\delta) = r_{\cos}(\delta),
\qquad
r_{\sin}(-\delta) = -r_{\sin}(\delta),
\]
showing that \(r_{\cos}\) is an even function and \(r_{\sin}\) is an odd function. This mirrors the symmetry of \(\cos(\delta)\) and \(\sin(\delta)\), respectively. Since AC power flow equations rely critically on how active and reactive power terms change sign when angle differences reverse direction, preserving this odd–even structure helps maintain the physical consistency of the model.

\subsubsection{Role of APF Parameter and Numerical Smoothness}
The parameter \(a\) controls the curvature of the all-pass functions as a function of the angle difference \(\delta\). If \(a\) is chosen too large, the functions become steep and may introduce numerical difficulties. If \(a\) is chosen too small, the approximation becomes overly flat and loses accuracy. Importantly, the denominator of the all-pass function never vanishes for real \(\delta\), so the model does not suffer from real-axis poles.

In practice, a moderate value such as \(a = 0.5\) provides a good balance between accuracy and numerical stability. With this choice, the APF remains smooth and well behaved over the range of angle differences typically encountered in AC OPF solutions, as confirmed by the numerical experiments reported later in the paper.

\subsubsection{Derivative Structure and Improved Conditioning}
Although the APF formulation remains nonlinear and nonconvex, its derivative structure differs from that of the classical trigonometric formulation. In the standard AC OPF model, the Jacobian and Hessian contain derivatives such as
\[
\frac{d}{d\delta}\sin(\delta) = \cos(\delta),
\qquad
\frac{d}{d\delta}\cos(\delta) = -\sin(\delta),
\]
which oscillate as \(\delta\) varies. These oscillations cause frequent sign changes and rapid curvature variation, which can lead to poorly conditioned Newton steps in interior-point methods.

In contrast, the derivatives of the all-pass components,
\[
\frac{d}{d\delta} r_{\cos}(\delta)
= \frac{-4a^2\delta}{(1 + (a\delta)^2)^2},
\qquad
\frac{d}{d\delta} r_{\sin}(\delta)
= \frac{2a\big(1 - (a\delta)^2\big)}{(1 + (a\delta)^2)^2},
\]
are smooth rational functions. They do not oscillate and exhibit gradual changes in curvature. As \(|\delta|\) increases, the denominator dominates and naturally damps the slope, leading to more stable and predictable Jacobian and Hessian matrices during optimization.

\subsubsection{Computational Efficiency of Rational Evaluations}
From a computational standpoint, the APF formulation replaces trigonometric evaluations with simple arithmetic operations involving addition, multiplication, and division. Both the functions and their derivatives can be computed efficiently without calling expensive trigonometric routines. For large-scale OPF problems, this reduction in per-iteration computational cost becomes significant, contributing to faster solver runtimes.

\subsubsection{Impact on Newton Steps and Solver Convergence}
\textcolor{black}{The smoother derivative structure of the APF OPF constraints may support better-conditioned Newton steps, with the intended benefit being improved computational efficiency rather than a greater likelihood of finding a better local optimum or avoiding suboptimal stationary points.}
In the numerical experiments, APF OPF achieves shorter solver times in many cases and fewer IPOPT iterations in a subset of cases. The objective and variable comparisons indicate comparable solution quality, with no claim of improved local optimality.

This effect becomes particularly pronounced in large-scale networks, where oscillatory trigonometric behavior can otherwise cause unstable or overly conservative step sizes.

Overall, the APF OPF formulation preserves the essential physics of PF, including loss modeling, angle symmetry, and directional behavior, while offering a numerical structure that is more compatible with interior-point optimization methods. By combining physical fidelity with smoother derivatives and reduced computational overhead, the APF formulation provides an accurate yet more solver-friendly alternative to the classical trigonometric AC OPF model.

\textcolor{black}{The preceding analysis extends to the
pre-rotated kernel used in the OPF formulation through
$\hat r(\delta_{ij})=e^{j\delta_{ij}^{dc}}r(\Delta_{ij})$,
with the DC reference held fixed during optimization.
The fixed rotation preserves unit magnitude, smoothness,
and the absence of real-axis poles. Since
$d^k\hat r/d\delta_{ij}^k
=e^{j\delta_{ij}^{dc}}r^{(k)}(\Delta_{ij})$,
it also preserves the complex derivative magnitudes at the
corresponding shifted argument. Symmetry is preserved as
conjugate symmetry when the angle difference and its DC
reference reverse sign together; the transformed components
are not generally even and odd about zero for a fixed nonzero
reference. The reference-dependent trigonometric coefficients
are computed once, so iterative evaluations remain rational.
These preserved kernel properties do not by themselves
guarantee improved conditioning of the complete OPF Newton
system; the computational benefits are assessed through
the numerical comparisons.}

\subsection{APF vs. Classical Kernels}
\label{subsec:allpass_vs_classic}

The functions $r_{\sin}(\delta)$ and $r_{\cos}(\delta)$ are plotted together with $\sin(\delta)$ and $\cos(\delta)$ over the interval $[-180^\circ, 180^\circ]$. These plots show the overall accuracy of the \textcolor{black}{rational} approximation and indicate the angle regions where the curves match more closely.

\begin{figure}[!t]
    \centering
    \captionsetup{font={footnotesize}}
    \begin{minipage}{0.48\columnwidth}
        \centering
        \includegraphics[width=\columnwidth]{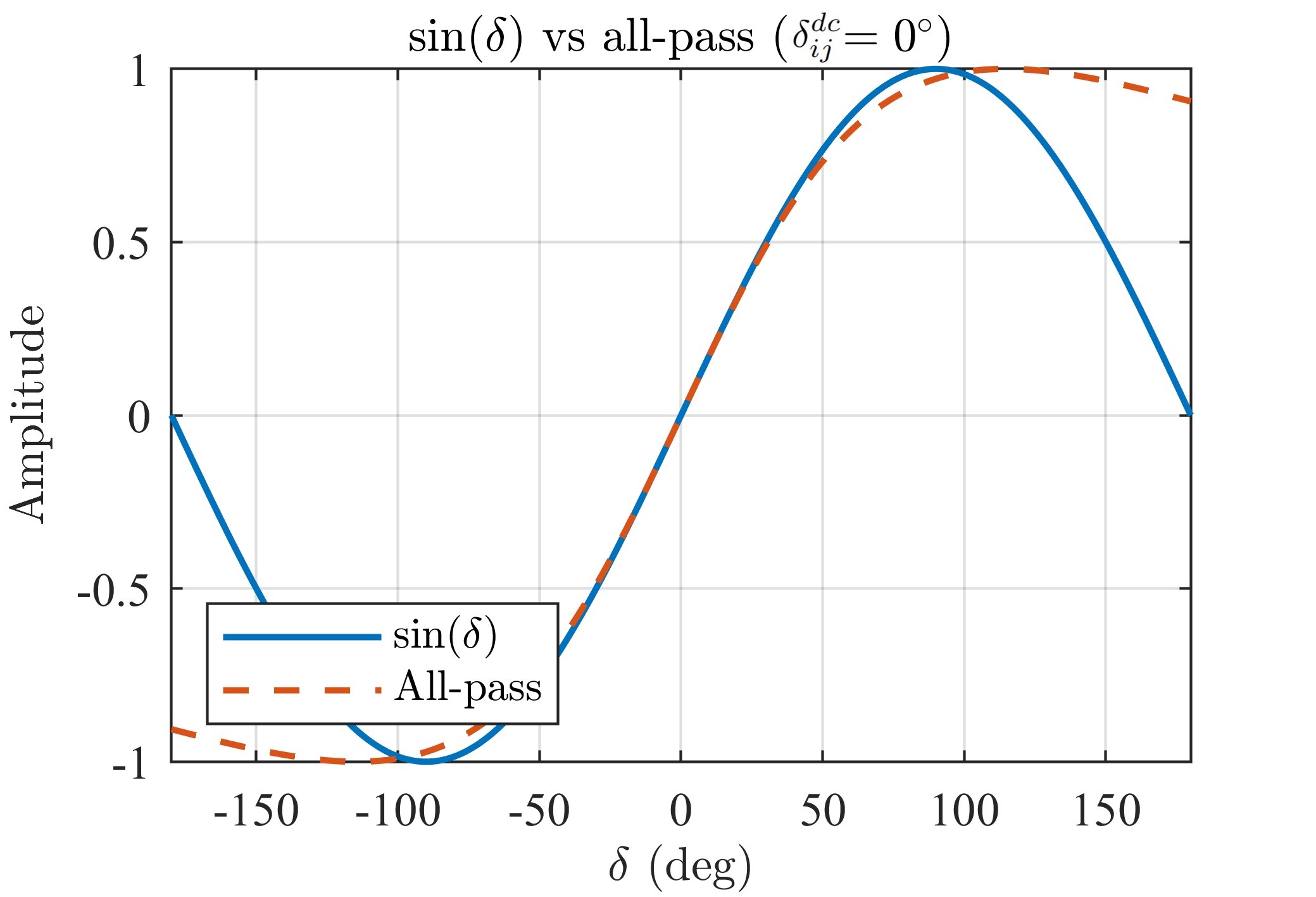}
        \caption*{(a) $\sin(\delta)$ vs $r_{\sin}(\delta)$}
        \label{fig:sin_vs_rsin}
    \end{minipage}
    \hfill
    \begin{minipage}{0.48\columnwidth}
        \centering
        \includegraphics[width=\columnwidth]{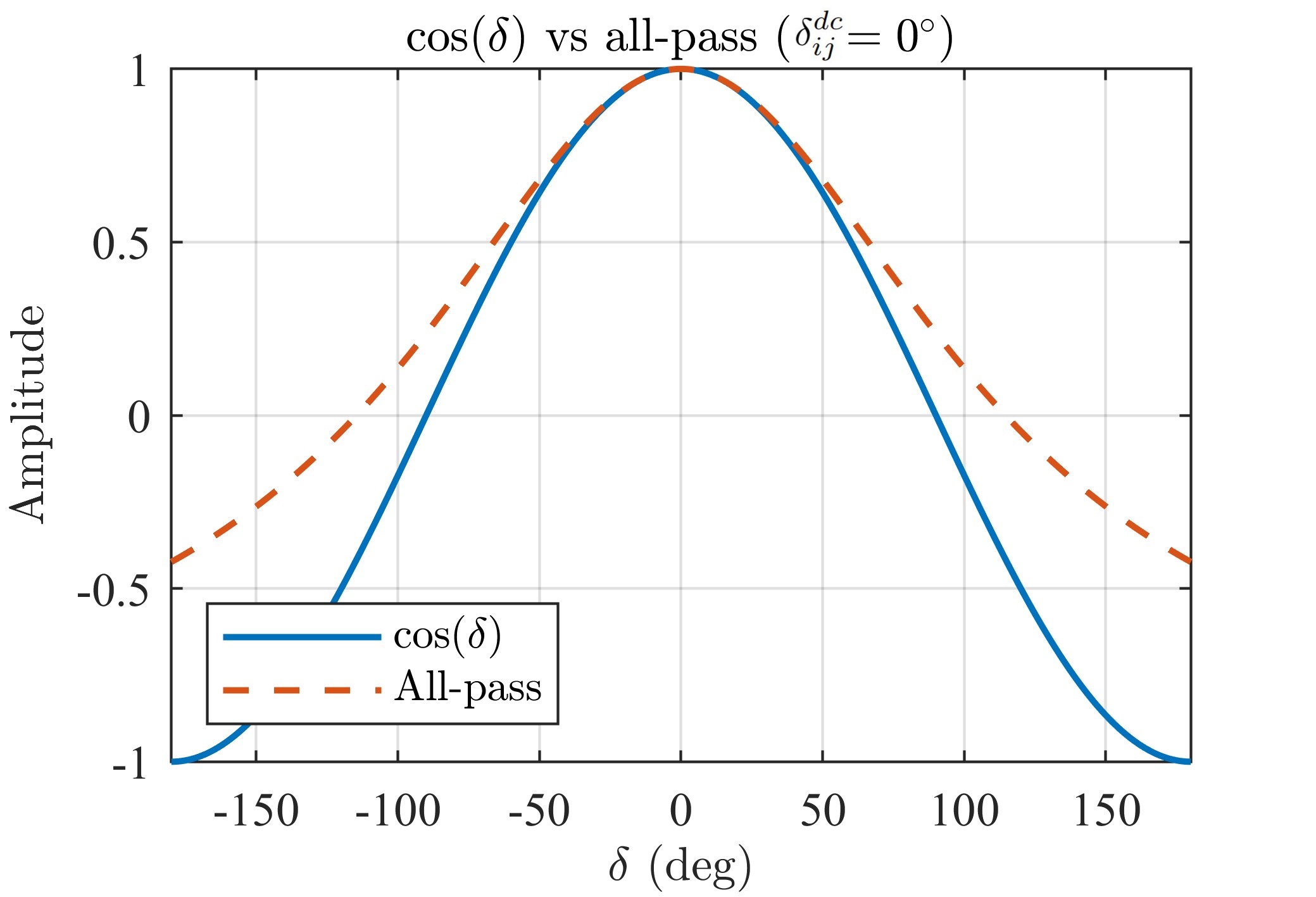}
        \caption*{(b) $\cos(\delta)$ vs $r_{\cos}(\delta)$}
        \label{fig:cos_vs_rcos}
    \end{minipage}

    \caption{Comparison of kernels}
    \label{fig:sin_cos_vs_allpass}
\end{figure}

Fig.~\ref{fig:sin_cos_vs_allpass}a shows $\sin(\delta)$ and $r_{\sin}(\delta)$ without any pre-rotation. Fig.~\ref{fig:sin_cos_vs_allpass}b shows the same comparison for the cosine function. In both cases, the all-pass functions follow the classical trigonometric curves well near the origin. Larger deviations appear when $|\delta|$ becomes large, which is outside typical operating ranges. This observation is consistent with how the first-order \textcolor{black}{rational} form is constructed, since it is centered at zero and provides its best accuracy near that point.

\begin{figure}[!t] \captionsetup{font={footnotesize}}
    \centering
    \includegraphics[width=0.65\columnwidth]{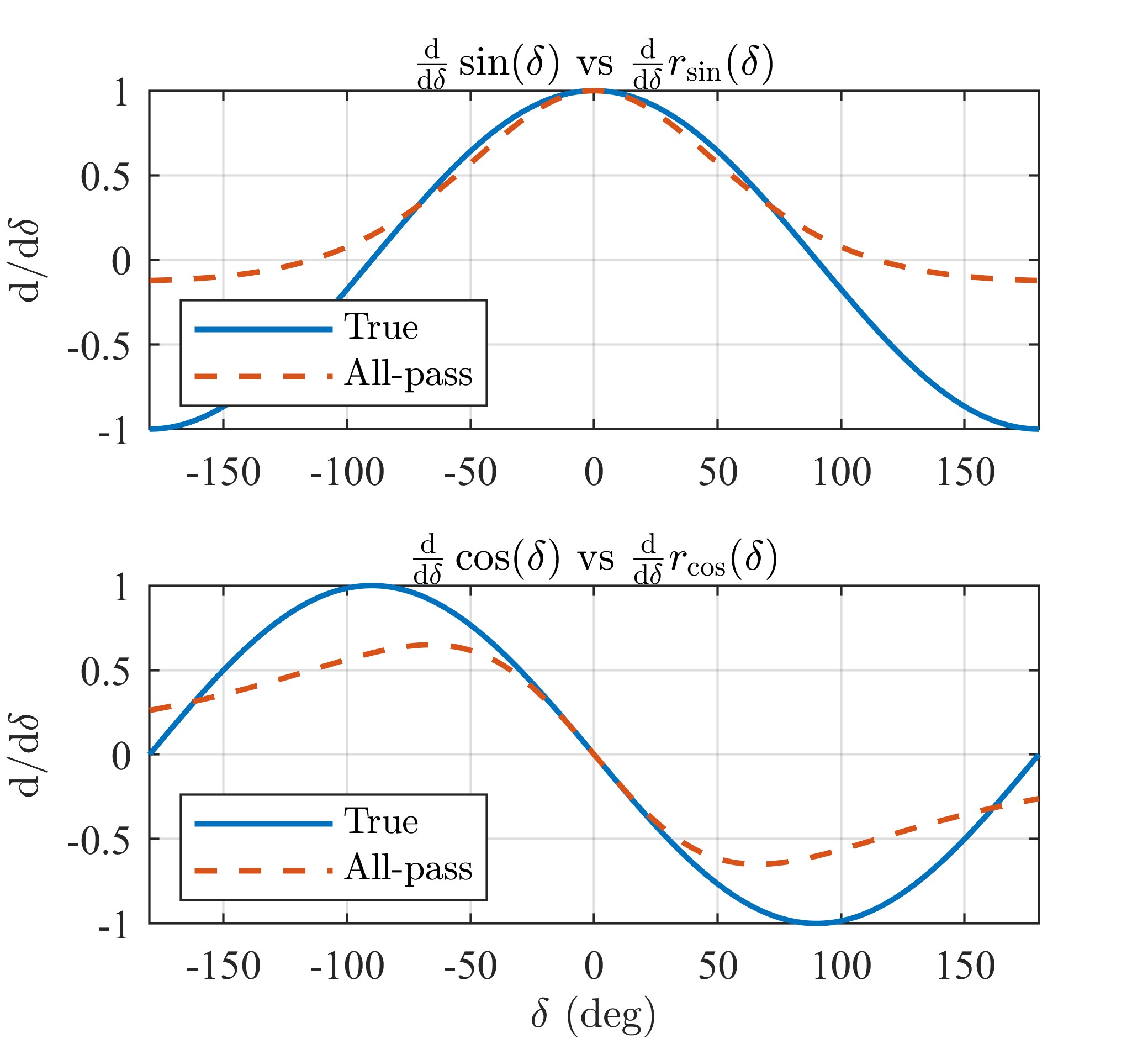}
    \caption{Comparison of the rate of variation}
    \label{fig:derivative_comparison}
\end{figure}

\begin{figure}[!t] \captionsetup{font={footnotesize}}
    \centering
    \begin{minipage}{0.45\columnwidth}
        \centering
        \includegraphics[width=\columnwidth]{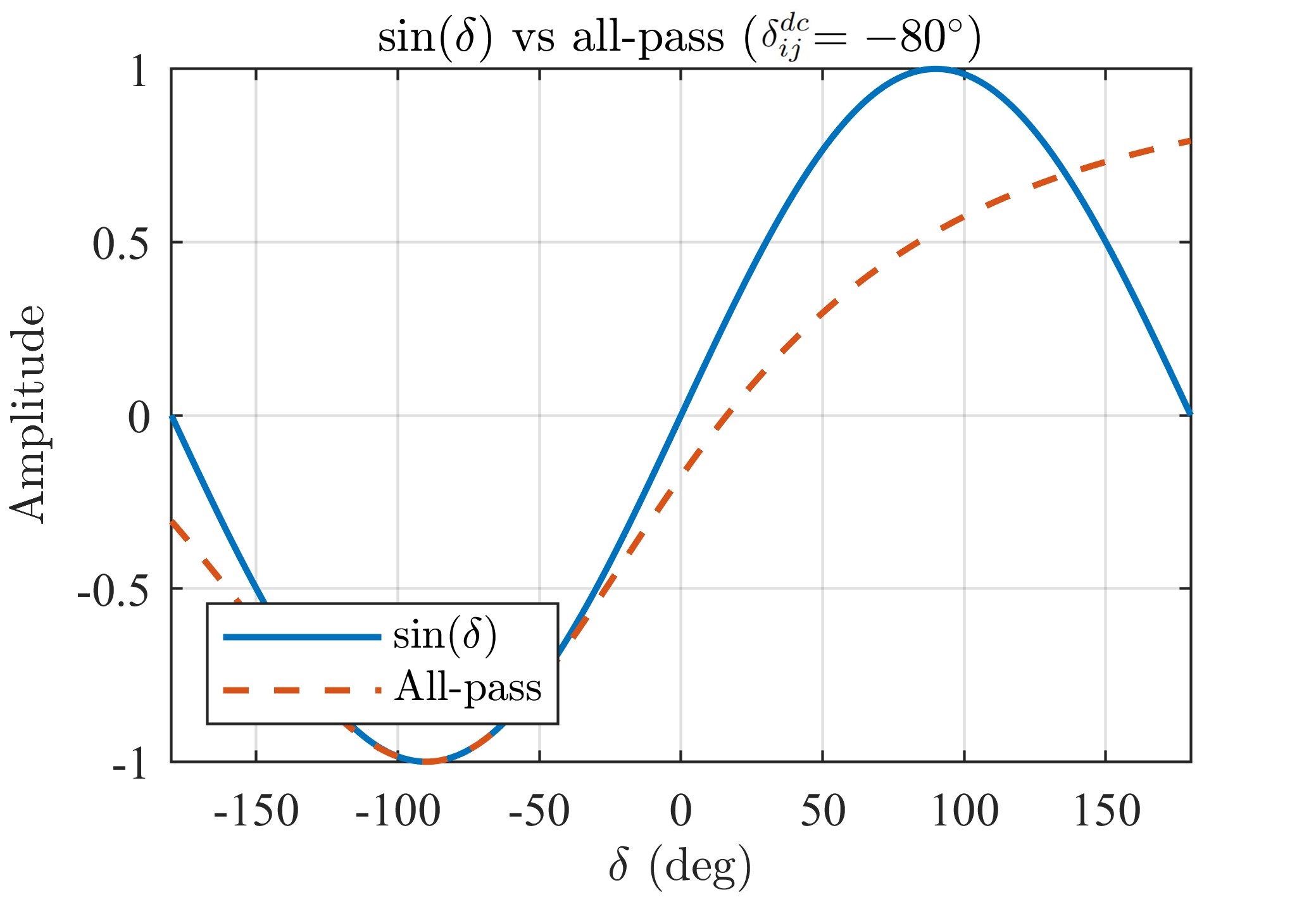}\\[1mm]
        \includegraphics[width=\columnwidth]{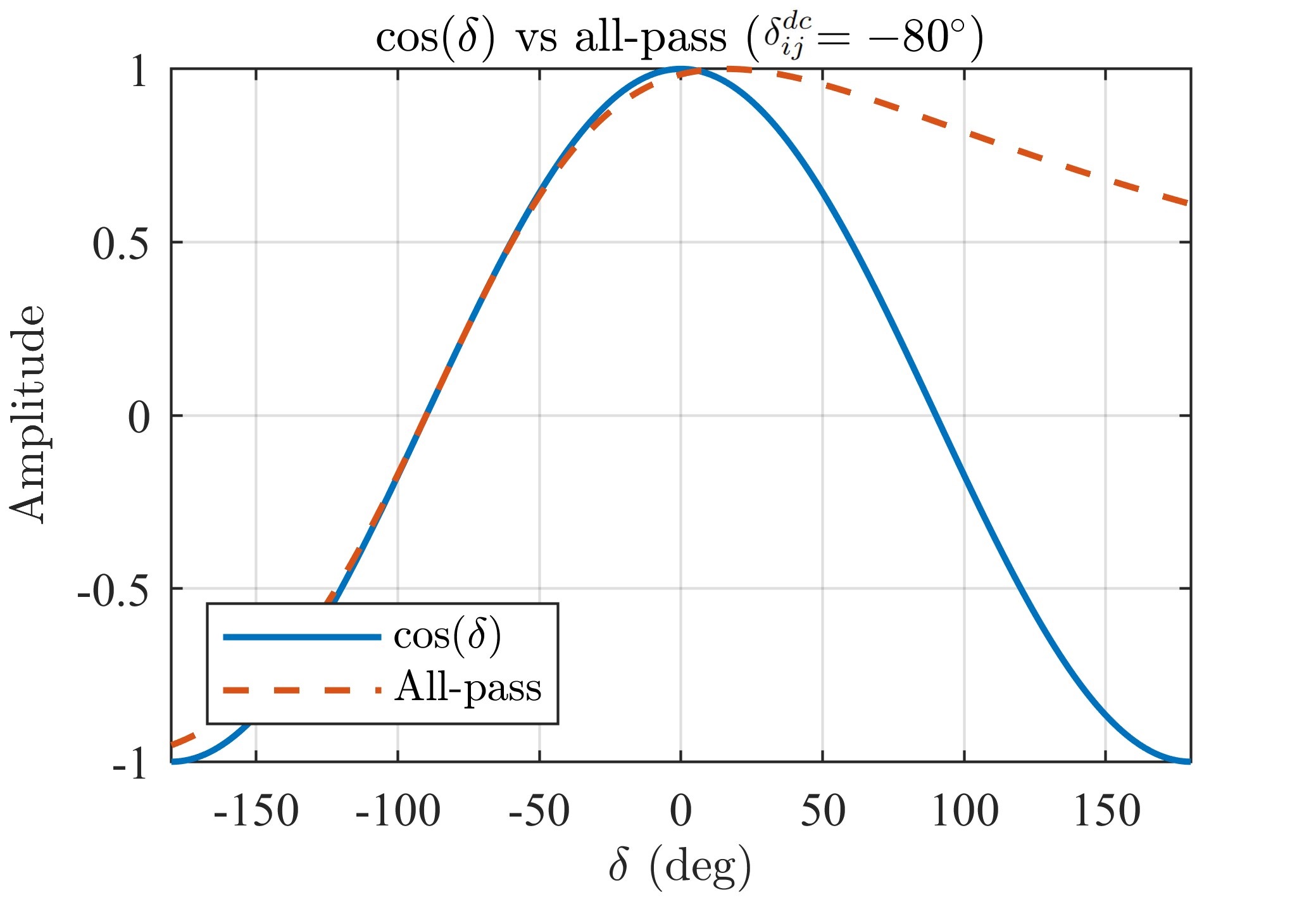}
    \end{minipage}
    \hfill
    \begin{minipage}{0.45\columnwidth}
        \centering
        \includegraphics[width=\columnwidth]{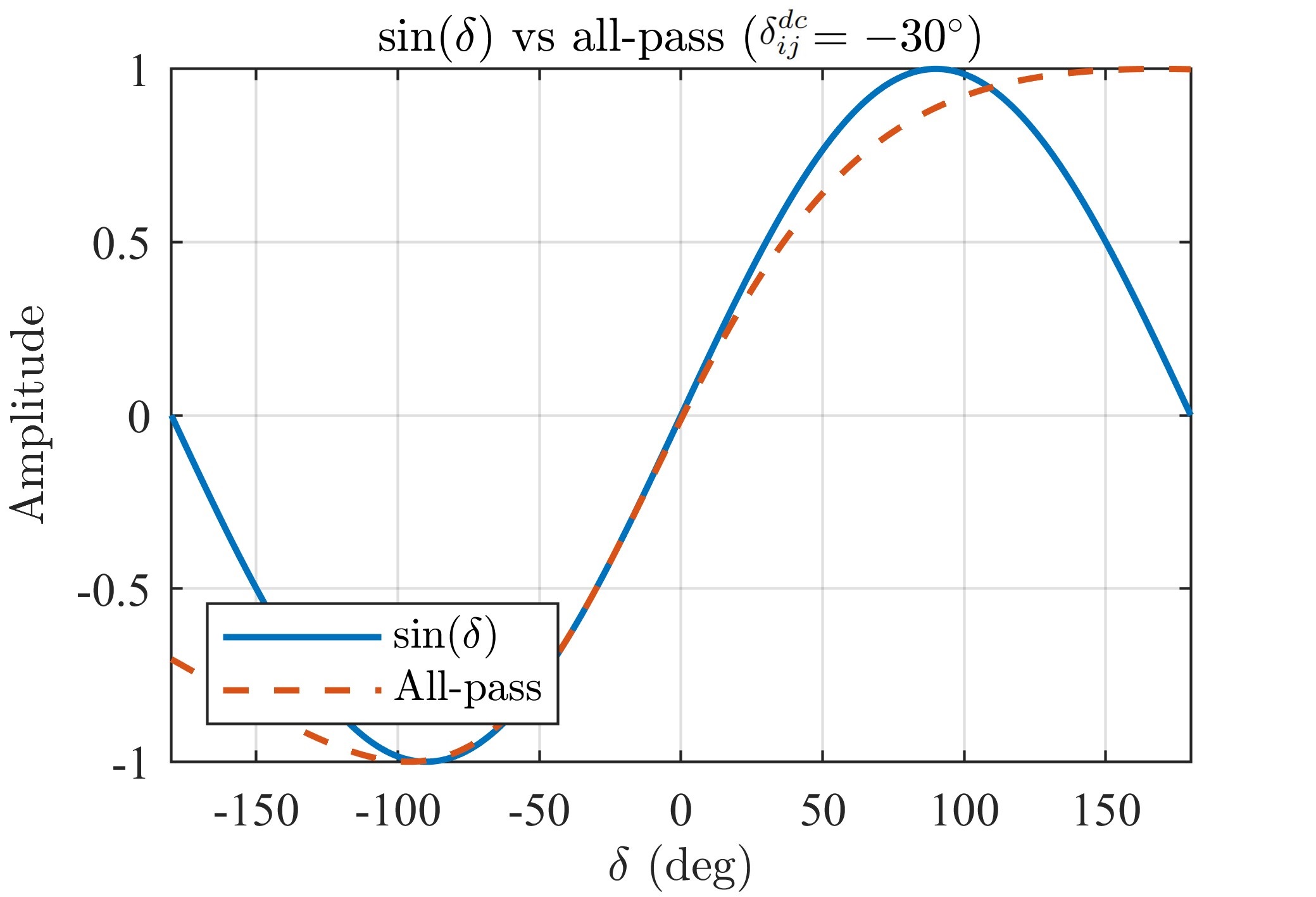}\\[1mm]
        \includegraphics[width=\columnwidth]{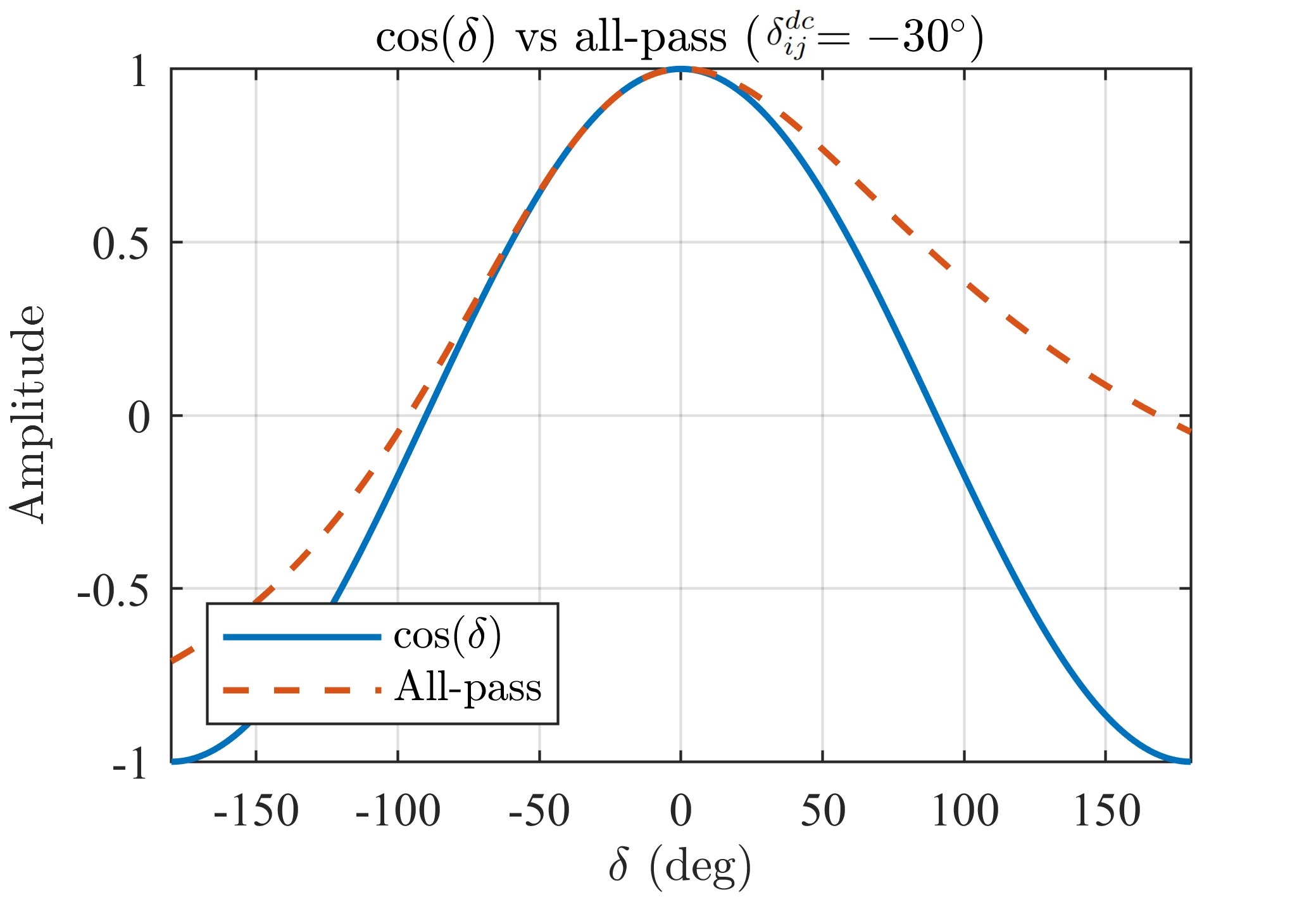}
    \end{minipage}

    \vspace{2mm}

    \begin{minipage}{0.45\columnwidth}
        \centering
        \includegraphics[width=\columnwidth]{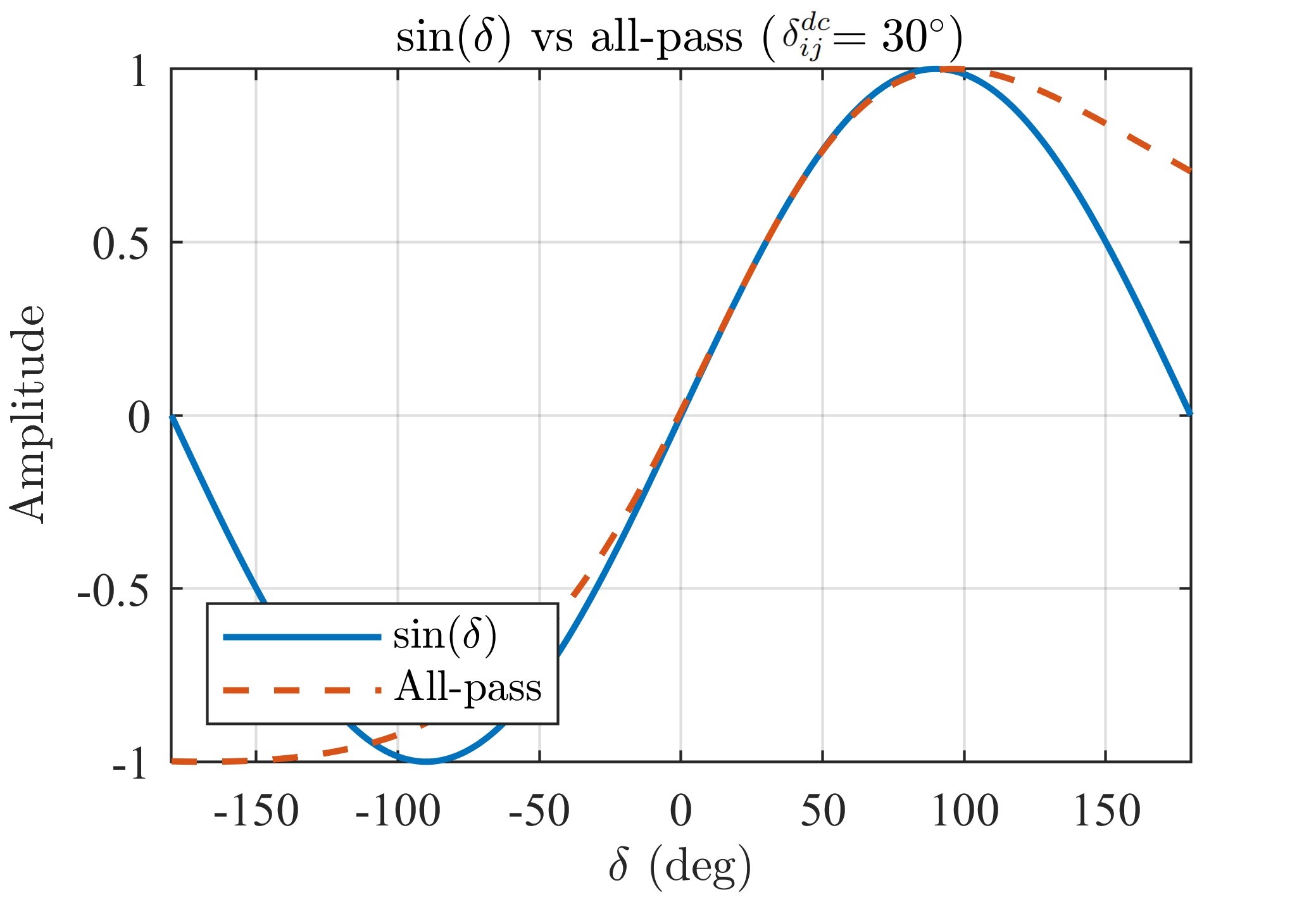}\\[1mm]
        \includegraphics[width=\columnwidth]{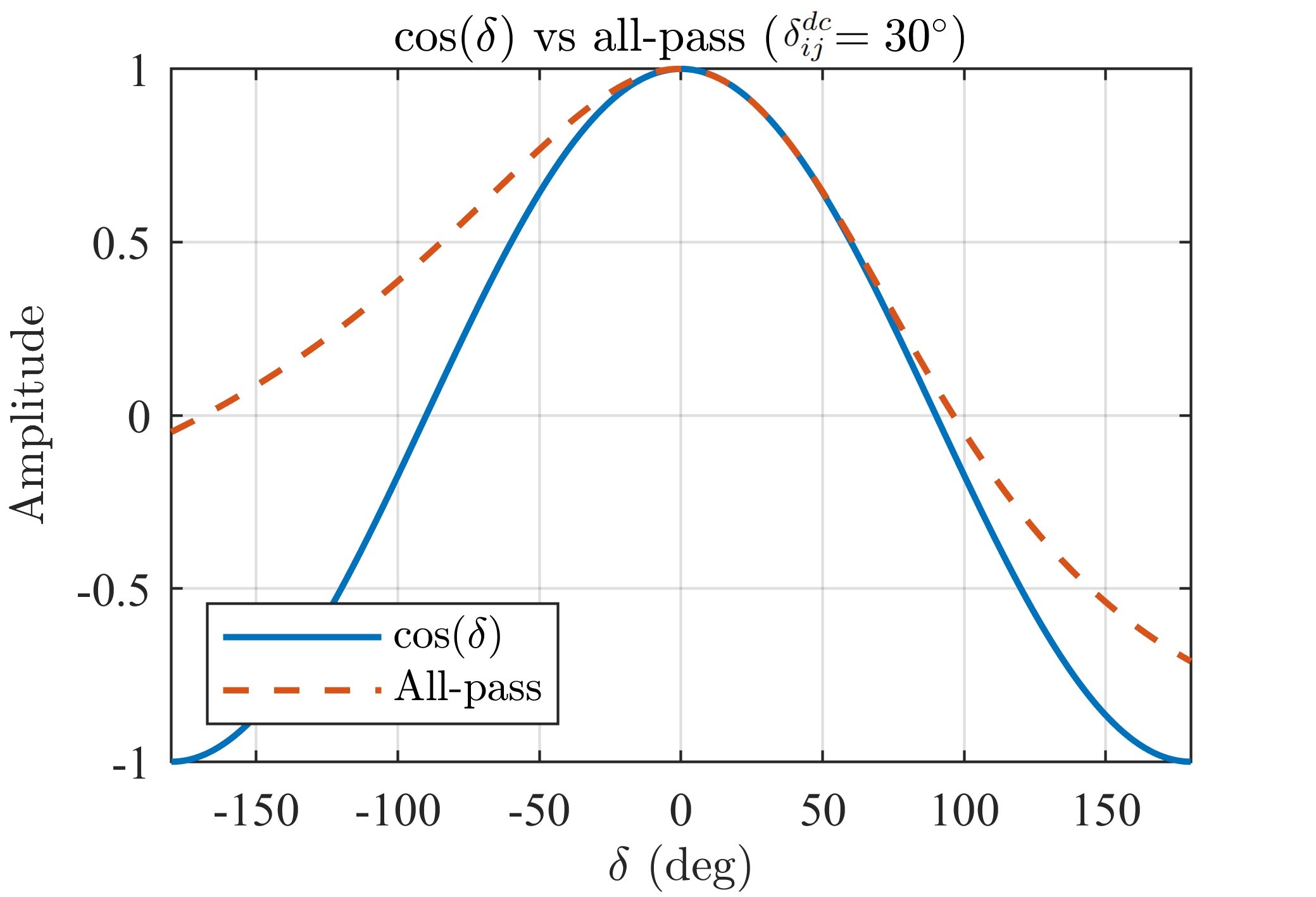}
    \end{minipage}
    \hfill
    \begin{minipage}{0.45\columnwidth}
        \centering
        \includegraphics[width=\columnwidth]{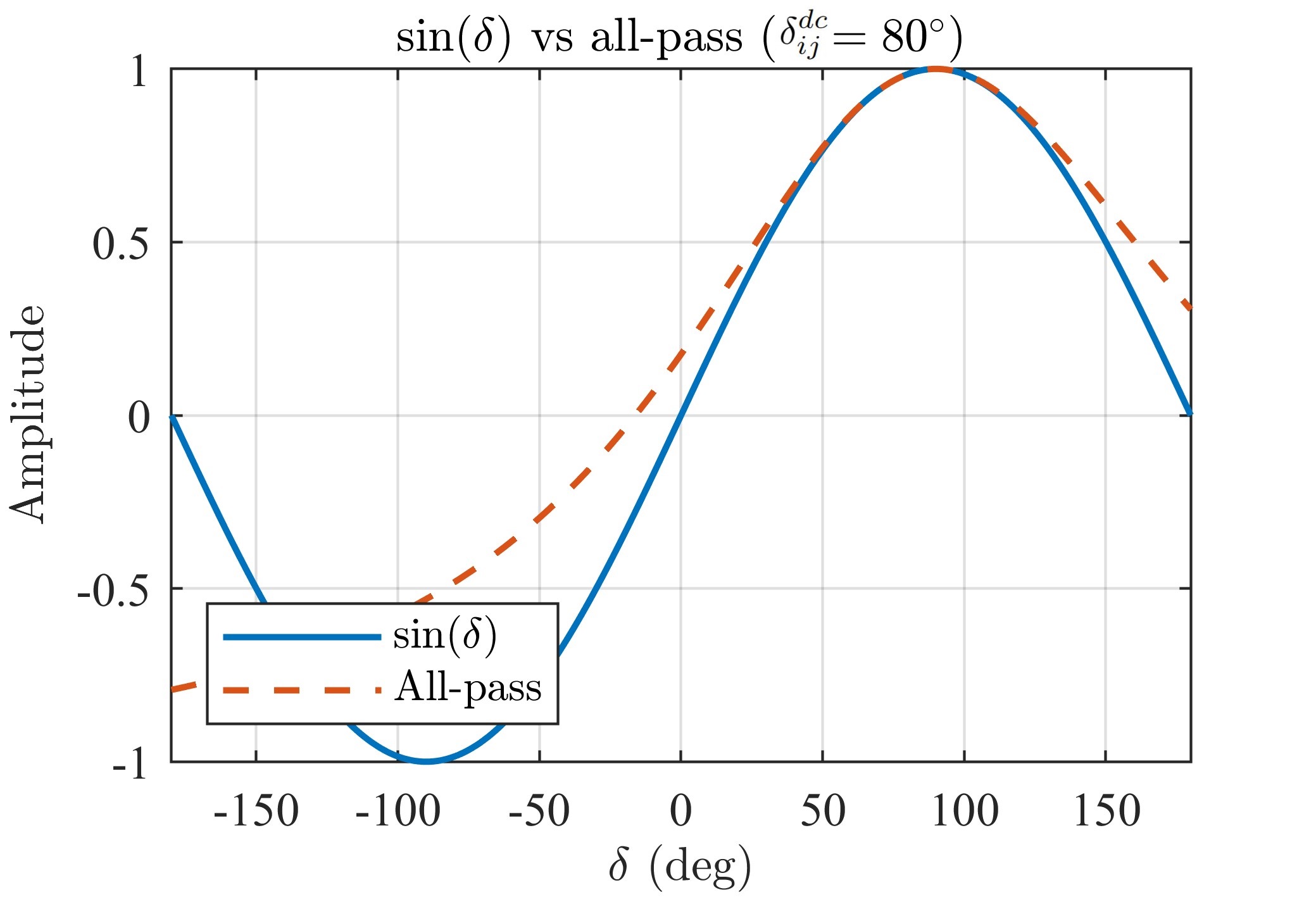}\\[1mm]
        \includegraphics[width=\columnwidth]{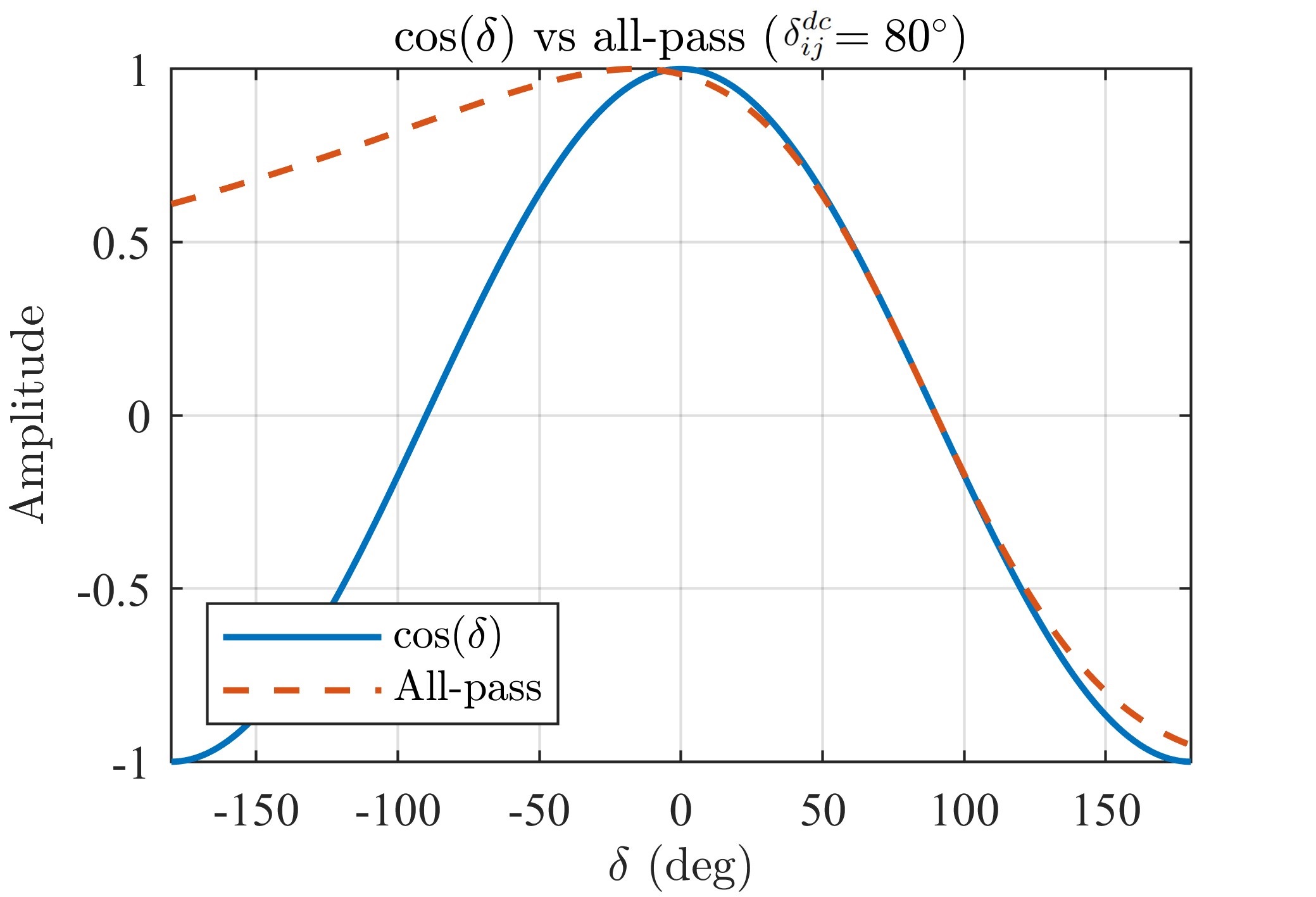}
    \end{minipage}
    \caption{Effect of applying different shifts to the argument of the all-pass kernels}
    \label{fig:pre_rotation_effects}
\end{figure}

The derivatives of the functions are also compared because they directly affect the Jacobian and Hessian used by IPM. The corresponding derivatives of the \textcolor{black}{rational} functions depend on smooth \textcolor{black}{rational} expressions whose slopes decrease gradually as $|\delta|$ increases. Fig.~\ref{fig:derivative_comparison} shows the comparison of the classical and all-pass derivatives over the same angle interval. Near zero, the slopes match closely, while for large angles the \textcolor{black}{rational} functions exhibit gentler behavior. \textcolor{black}{The shift used in Figs.~\ref{fig:sin_cos_vs_allpass}
and~\ref{fig:pre_rotation_effects} is denoted by
$\delta_{ij}^{dc}$, consistently with the DC reference angle
difference defined in~\eqref{eq:delta_dc}.}

With the pre-rotation applied to the APF OPF approximation, the argument of the all-pass functions becomes $\Delta_{ij}$, which is close to zero for many lines. To show how shifting the argument affects the approximation, several plots with different shift values are provided in Fig.~\ref{fig:pre_rotation_effects}. In each case, $r_{\sin}(\delta - \text{shift})$ and $r_{\cos}(\delta - \text{shift})$ are compared with the corresponding trigonometric functions. The location where the curves align most closely moves horizontally with the shift. This illustrates how the shift places the high-accuracy window of the \textcolor{black}{rational} functions near the actual operating angles of the network.

These comparisons indicate that APF closely follows the classical trigonometric kernels in the operating region of OPF and that pre-rotation further improves the alignment by centering the approximation around expected angle differences.

\section{Numerical Experiments and Results}\label{sec:numerical}
We evaluate the numerical performance and solution quality of the proposed APF OPF formulation in \eqref{eq:apf_opf} relative to the classical AC OPF model in \eqref{eq:acopf_exp}. All simulations are reproducible using the publicly available repository at \href{https://github.com/LSU-RAISE-LAB/APF-OPF}{GitHub}, which includes the codebase, configuration files, and scripts for regenerating the reported figures and tables.

\subsection{Numerical Setup and Solvers}
All simulations are carried out using IPOPT \cite{wachter2006implementation}, which is an interior-point optimizer designed for large nonlinear problems. IPOPT solves a sequence of barrier subproblems and relies on Newton steps, so its performance is sensitive to the smoothness and well-conditioning of the model equations. This makes it an appropriate platform for evaluating whether replacing trigonometric terms with the APF kernel produces more stable numerical behavior.

The AC OPF \eqref{eq:acopf_exp} and APF OPF \eqref{eq:apf_opf} are implemented in two different environments: YALMIP \cite{lofberg2004yalmip} and CasADi \cite{andersson2019casadi}. YALMIP provides a symbolic MATLAB interface that is easy to develop and debug, while CasADi builds a computational graph and generates efficient derivative code. For small and medium-sized networks, both work well, but CasADi generally scales better on large PGLib cases due to its optimized automatic differentiation and sparsity handling. In total, four solver functions are used:
\begin{itemize}
    \item \texttt{acopf\_exp} (YALMIP + IPOPT, AC OPF),
    \item \texttt{acopf\_allpass} (YALMIP + IPOPT, APF OPF),
    \item \texttt{acopf\_exp\_ipopt\_casadi} (CasADi + IPOPT, AC OPF),
    \item \texttt{acopf\_allpass\_ipopt\_casadi} (CasADi + IPOPT, APF OPF).
\end{itemize}

All four functions accept the same core inputs: a MATPOWER case name, a percentage parameter \(m\) for optional scaling of line ratings (for manually making 90 percent of lines have reduced \(m\) percent line rates), and an optional IPOPT override structure. The APF versions accept an additional vector \(a_{\text{params}}\) for specifying the poles of the \textcolor{black}{rational} kernel; if omitted, a single pole is chosen automatically from the network’s angle band. Each function returns a result structure containing bus voltages, generator outputs, objective value, angle information, congestion, solver time, and (for the APF solvers) the pre-rotation time and the edge-angle differences.

To evaluate the accuracy, optimality, and physical feasibility of the proposed all-pass formulation, a dedicated comparison routine,
\texttt{compare\_acopf\_results(res\_exp, ok\_exp, res\_allpass, ok\_allpass)}, is employed. This routine preserves both solver outputs intact and compares them in a standardized manner. Optimality is assessed via absolute and relative objective gaps between the AC OPF and APF OPF solutions. Angle behavior is analyzed directly from solver outputs using raw bus angle differences and kernel-level edge angle differences, for which maximum, average, and minimum values are reported.

In this context, feasibility refers to whether the APF OPF solution, when evaluated outside the optimization model, satisfies the original nonlinear AC power flow equations and all operational constraints. Specifically, the APF OPF voltage magnitudes, phase angles, and generator outputs are substituted into the exact trigonometric AC equations to recompute nodal active and reactive power injections and branch power flows. These recomputed quantities are then checked against power-balance equations, voltage magnitude bounds, generator active and reactive limits, branch thermal limits, and branch angle-difference limits. All feasibility checks are performed using a user-defined numerical tolerance, which accounts for floating-point errors and solver termination accuracy; a constraint is considered satisfied if its violation does not exceed this tolerance. Feasibility is reported both as an overall pass/fail indicator and through detailed violation counts that identify the number and location of buses, generators, and branches violating each constraint class.

Congestion is evaluated using solver-reported binding constraints, and congested lines are compared between the AC OPF and APF OPF solutions using physical (undirected) line matching. \textcolor{black}{Reported timings exclude model construction,
parsing, and compilation in YALMIP/CasADi. APF OPF timings in
Tables~\ref{tab:allpass_global_summary} and
\ref{tab:iteration_time_reductions} include solver execution
and pre-rotation, whereas Tables~\ref{tab:load_variation} and
\ref{tab:a_sensitivity} report solver execution.}

All experiments are executed through a main driver script, \texttt{run\_acopf\_compare(case\_name, mode, m)}, where \texttt{mode} selects YALMIP or CasADi. The script automatically calls both AC OPF and APF OPF solvers, collects their results, runs the comparison function, and prints the numerical report. This creates a unified and repeatable numerical experiment pipeline for all IEEE and PGLib test cases used in the paper. 

\subsection{Benchmark Results Across Network Sizes}
The test set spans from small IEEE and MATPOWER networks to large-scale PGLib transmission systems with more than 10{,}000 buses. All experiments are conducted using identical solver configurations to isolate the impact of the kernel approximation from solver parameter tuning.

Table~\ref{tab:allpass_global_summary} reports unified performance and diagnostic metrics for all benchmark cases, including time speedup, optimality gap, feasibility under true AC re-evaluation, detailed counts of constraint violations, congestion consistency, and the maximum branch-wise voltage-angle difference $\theta_i-\theta_j$. For clarity, the results are discussed according to three size categories: small systems (9--300 buses), medium systems (588--3{,}375 buses), and large systems (4{,}601--10{,}480 buses).

We first clarify how the feasibility-related quantities reported in Table~\ref{tab:allpass_global_summary} are computed. Feasibility is evaluated by substituting the APF OPF solution into the original nonlinear AC power flow equations and operational constraints. In particular, bus active and reactive power balance mismatches are computed by replacing the APF OPF voltage magnitudes, phase angles, and generator outputs into the exact trigonometric AC power flow equations, and the resulting mismatches are explicitly evaluated and reported.

All remaining feasibility checks correspond to inequality constraints, including voltage magnitude limits, generator active and reactive power limits, branch thermal limits, and branch angle-difference limits. For these constraints, a marginal mismatch metric is reported. Specifically, a value of zero indicates that the constraint is satisfied for all elements, while any nonzero value represents the magnitude by which a constraint is violated. For each constraint class, Table~\ref{tab:allpass_global_summary} reports the number of elements whose violation exceeds a specified numerical tolerance, along with the maximum and average violation magnitudes across all elements. These columns therefore provide a direct assessment of whether the APF OPF solution is physically feasible when evaluated under the exact AC OPF constraints.

In addition to feasibility diagnostics, Table~\ref{tab:mismatch_summary_med_large} reports variable mismatches between the AC OPF and APF OPF solutions. In this table, mismatches in generator dispatch, voltage magnitudes, voltage angles, and branch power flows are computed by directly comparing the outputs of the two formulations. Only the maximum and average mismatch values are reported. This analysis is limited to medium- and large-scale systems, as mismatches in small-scale networks are consistently negligible and do not provide additional insight.

A few large-scale cases, such as \texttt{case8387\_pegase}, exhibit feasibility or congestion mismatches. These cases correspond to networks where the classical AC OPF model itself converges only to acceptable-level solutions or approaches step-length failure, making both formulations sensitive to numerical conditioning. In all well-conditioned networks, the AC OPF and APF OPF formulations agree in feasibility outcomes, congestion patterns, and voltage-angle behavior.

Across all benchmark cases, three formulations are evaluated: the classical AC OPF, the APF OPF with DC power flow based pre-rotation, and the APF OPF with DC OPF based pre-rotation. Numerical experiments indicate that DC-OPF--based pre-rotation generally yields slightly more accurate solutions than DC-PF--based pre-rotation when compared against the classical AC OPF. Nevertheless, the solutions obtained using both pre-rotation strategies remain highly comparable to the AC OPF in terms of objective value, feasibility, congestion patterns, and voltage-angle behavior.

Despite its improved accuracy, DC-OPF--based pre-rotation often results in longer solver times compared to both the classical AC OPF and the APF OPF with DC-PF--based pre-rotation. In contrast, DC-PF--based pre-rotation consistently achieves faster solution times across all tested cases while maintaining solution accuracy that is nearly indistinguishable from the classical AC OPF. For this reason, DC-PF--based pre-rotation is adopted as the default configuration throughout the numerical study.

Accordingly, all results reported in Tables~\ref{tab:allpass_global_summary} and \ref{tab:mismatch_summary_med_large} are obtained using DC-PF--based pre-rotation, with the exception of \texttt{case118}, \texttt{case8387\_pegase}, \texttt{case9591\_goc}, and \texttt{case10480\_goc}. In these cases, DC-OPF--based pre-rotation provides more accurate solutions while also yielding smaller solver times than DC-PF--based pre-rotation, and is therefore used in place of the default setting.

\begin{table*}[!t]
\footnotesize
\setlength{\tabcolsep}{2.4pt}
\renewcommand{\arraystretch}{1.08}
\captionsetup{font={footnotesize}}
\caption{Summary of solver performance across benchmark cases with feasibility and congestion diagnostics}
\centering
\newcolumntype{T}{@{}c c c@{}}

\resizebox{\textwidth}{!}{%
\begin{tabular}{l |
c |
c |
T @{\hspace{6pt}}
T @{\hspace{6pt}}
T @{\hspace{6pt}}
T @{\hspace{6pt}}
T @{\hspace{6pt}}
T @{\hspace{6pt}}
T |
@{}c@{\hspace{4pt}}c@{\hspace{4pt}}c@{} |
c}
\toprule
\rowcolor{headergray}
\textbf{Case} &
\textbf{Time} &
\textbf{Opt.} &
\multicolumn{21}{c|}{\textbf{Feasibility Check with Constraint Violations (Tolerance)}} &
\multicolumn{3}{c|}{\textbf{Congestion}} &
\textbf{Max} \\
\cmidrule(lr){4-24}\cmidrule(lr){25-27}

\rowcolor{headergray}&
\textbf{Spd.} &
\textbf{Gap} &
\multicolumn{3}{c}{\textbf{Bus Bal. P (e-1)}} &
\multicolumn{3}{c}{\textbf{Bus Bal. Q (e-1)}} &
\multicolumn{3}{c}{\textbf{$V$ (e-4)}} &
\multicolumn{3}{c}{\textbf{$p^g$ (e-2)}} &
\multicolumn{3}{c}{\textbf{$q^g$ (e-2)}} &
\multicolumn{3}{c}{\textbf{$\theta_i-\theta_j$ (e-3)}} &
\multicolumn{3}{c|}{\textbf{$S_{ij}$ (e-2)}} &
\textbf{AC} & \textbf{APF} & \textbf{Mis} &
$\theta_i-\theta_j$ \\
\cmidrule(lr){4-6}\cmidrule(lr){7-9}\cmidrule(lr){10-12}
\cmidrule(lr){13-15}\cmidrule(lr){16-18}\cmidrule(lr){19-21}
\cmidrule(lr){22-24}

\rowcolor{headergray}&
(\%) &
(\%) &
\textbf{\#} & \textbf{Max} & \textbf{Avg} &
\textbf{\#} & \textbf{Max} & \textbf{Avg} &
\textbf{\#} & \textbf{Max} & \textbf{Avg} &
\textbf{\#} & \textbf{Max} & \textbf{Avg} &
\textbf{\#} & \textbf{Max} & \textbf{Avg} &
\textbf{\#} & \textbf{Max} & \textbf{Avg} &
\textbf{\#} & \textbf{Max} & \textbf{Avg} &
\# & \# & \# &
(deg) \\
\midrule
\midrule

\rowcolor{smallscale}
\multicolumn{28}{c}{\textbf{Small--Scale Systems (9--300 buses)}} \\
\midrule

\texttt{case9} & 45 & 0.0000 &
\cmark & 5e-5 & 1e-5 &
\cmark & 1e-5 & 3e-6 &
\cmark & 1e-8 & 4e-9 &
\cmark & 0 & 0 &
\cmark & 0 & 0 &
\cmark & 0 & 0 &
 & 0 & 0 &
0 & 0 & 0 & 5.5 \\

\texttt{case12da} & 22 & 0.0000 &
\cmark & 3e-7 & 9e-8 &
\cmark & 8e-7 & 2e-7 &
\cmark & 0 & 0 &
\cmark & 0 & 0 &
\cmark & 0 & 0 &
\cmark & 0 & 0 &
\cmark & 0 & 0 &
0 & 0 & 0 & 0.3 \\

\texttt{case18} & 16 & 0.0000 &
\cmark & 6e-7 & 7e-8 &
\cmark & 2e-7 & 4e-8 &
\cmark & 0 & 0 &
\cmark & 0 & 0 &
\cmark & 0 & 0 &
\cmark & 0 & 0 &
\cmark & 0 & 0 &
0 & 0 & 0 & 4.2 \\

\texttt{case22} & 17 & 0.0000 &
\cmark & 1e-7 & 3e-8 &
\cmark & 2e-7 & 5e-8 &
\cmark & 0 & 0 &
\cmark & 0 & 0 &
\cmark & 0 & 0 &
\cmark & 0 & 0 &
\cmark & 0 & 0 &
0 & 0 & 0 & 0.1 \\

\texttt{case24\_ieee\_rts} & 38 & 0.0000 &
\cmark & 6e-5 & 8e-6 &
\cmark & 1e-5 & 2e-6 &
\cmark & 1e-8 & 2e-9 &
\cmark & 4e-8 & 1e-8 &
\cmark & 1e-8 & 2e-9 &
\cmark & 0 & 0 &
\cmark & 0 & 0 &
0 & 0 & 0 & 11.6 \\

\texttt{case30} & 11 & 0.0000 &
\cmark & 2e-5 & 2e-6 &
\cmark & 7e-7 & 1e-7 &
\cmark & 1e-8 & 4e-10 &
\cmark & 0 & 0 &
\cmark & 0 & 0 &
\cmark & 0 & 0 &
\cmark & 2e-7 & 4e-9 &
2 & 2 & 0 & 2.5 \\

\texttt{case33bw} & 24 & 0.0000 &
\cmark & 3e-7 & 4e-8 &
\cmark & 6e-7 & 6e-8 &
\cmark & 0 & 0 &
\cmark & 0 & 0 &
\cmark & 0 & 0 &
\cmark & 0 & 0 &
\cmark & 0 & 0 &
0 & 0 & 0 & 0.2 \\

\texttt{case38si} & 28 & 0.0000 &
\cmark & 3e-6 & 4e-7 &
\cmark & 6e-6 & 5e-7 &
\cmark & 0 & 0 &
\cmark & 0 & 0 &
\cmark & 0 & 0 &
\cmark & 0 & 0 &
\cmark & 0 & 0 &
0 & 0 & 0 & 0.2 \\

\texttt{case39} & 45 & 0.0000 &
\cmark & 2e-3 & 1e-4 &
\cmark & 2e-4 & 2e-5 &
\cmark & 1e-8 & 1e-9 &
\cmark & 7e-8 & 3e-8 &
\cmark & 3e-8 & 7e-9 &
\cmark & 0 & 0 &
\cmark & 0 & 0 &
0 & 0 & 0 & 6.7 \\

\texttt{case51ga} & 28 & 0.0000 &
\cmark & 7e-7 & 8e-8 &
\cmark & 9e-7 & 1e-7 &
\cmark & 0 & 0 &
\cmark & 0 & 0 &
\cmark & 0 & 0 &
\cmark & 0 & 0 &
\cmark & 0 & 0 &
0 & 0 & 0 & 0.2 \\

\texttt{case57} & 72 & 0.0000 &
\cmark & 1e-3 & 4e-5 &
\cmark & 2e-4 & 1e-5 &
\cmark & 1e-8 & 2e-10 &
\cmark & 0 & 0 &
\cmark & 1e-8 & 3e-9 &
\cmark & 0 & 0 &
\cmark & 0 & 0 &
0 & 0 & 0 & 4.9 \\

\texttt{case69} & 31 & 0.0000 &
\cmark & 4e-7 & 2e-8 &
\cmark & 1e-6 & 5e-8 &
\cmark & 0 & 0 &
\cmark & 0 & 0 &
\cmark & 0 & 0 &
\cmark & 0 & 0 &
\cmark & 0 & 0 &
0 & 0 & 0 & 0.4 \\

\texttt{case74ds} & 36 & 0.0000 &
\cmark & 9e-7 & 3e-8 &
\cmark & 1e-6 & 3e-8 &
\cmark & 0 & 0 &
\cmark & 0 & 0 &
\cmark & 0 & 0 &
\cmark & 0 & 0 &
\cmark & 0 & 0 &
0 & 0 & 0 & 0.02 \\

\texttt{case118} & 47 & 0.0000 &
\cmark & 3e-4 & 1e-5 &
\cmark & 9e-5 & 3e-6 &
\cmark & 1e-8 & 8e-10 &
\cmark & 1e-8 & 3e-9 &
\cmark & 1e-8 & 2e-9 &
\cmark & 0 & 0 &
\cmark & 0 & 0 &
0 & 0 & 0 & 10.7 \\

\texttt{case141} & 27 & 0.0000 &
\cmark & 2e-6 & 3e-8 &
\cmark & 3e-6 & 5e-8 &
\cmark & 0 & 0 &
\cmark & 0 & 0 &
\cmark & 0 & 0 &
\cmark & 0 & 0 &
\cmark & 0 & 0 &
0 & 0 & 0 & 0.1 \\

\texttt{case300} & 38 & 0.0000 &
\cmark & 4e-3 & 6e-5 &
\cmark & 4e-4 & 7e-6 &
\cmark & 1e-8 & 1e-9 &
\cmark & 1e-8 & 6e-10 &
\cmark & 4e-8 & 4e-9 &
\cmark & 0 & 0 &
\cmark & 0 & 0 &
0 & 0 & 0 & 20.7 \\

\midrule
\rowcolor{mediumscale}
\multicolumn{28}{c}{\textbf{Medium--Scale \texttt{pglib\_opf} Systems (588--3375 buses)}} \\
\midrule

\texttt{ case588\_sdet} & 20 & 0.0018 &
\cmark & 4e-2 & 8e-4 &
\cmark & 1e-2 & 1e-4 &
\cmark & 1e-8 & 9e-11 &
\cmark & 1e-7 & 2e-8 &
\cmark & 4e-8 & 2e-9 &
\cmark & 0 & 0 &
\cmark & 2e-3 & 4e-6 &
15 & 15 & 0 & 11.0 \\

\texttt{ case1354\_pegase} & 27 & 0.0001 &
\cmark & 2e-2 & 1e-4 &
\cmark & 1e-3 & 2e-5 &
\cmark & 1e-8 & 3e-10 &
\cmark & 4e-7 & 3e-8 &
\cmark & 4e-8 & 4e-9 &
\cmark & 0 & 0 &
\cmark & 3e-4 & 3e-7 &
15 & 15 & 0 & 13.1 \\

\texttt{ case2383wp\_k} & 50 & 0.0001 &
\cmark & 5e-2 & 5e-5 &
\cmark & 3e-3 & 5e-6 &
\cmark & 1e-8 & 1e-10 &
\cmark & 1e-7 & 1e-8 &
\cmark & 2e-8 & 4e-9 &
\cmark & 0 & 0 &
\cmark & 1e-3 & 4e-7 &
5 & 5 & 0 & 13.5 \\

\texttt{ case2746wp\_k} & 31 & 0.0000 &
1 & 3e-1 & 1e-3 &
\cmark & 2e-2 & 4e-4 &
\cmark & 1e-8 & 7e-11 &
\cmark & 4e-8 & 2e-9 &
\cmark & 0 & 0 &
\cmark & 0 & 0 &
\cmark & 0 & 0 &
0 & 0 & 0 & 13.9 \\

\texttt{ case2869\_pegase} & 40 & 0.0000 &
\cmark & 4e-2 & 1e-4 &
\cmark & 2e-3 & 1e-5 &
\cmark & 1e-8 & 1e-10 &
\cmark & 4e-7 & 2e-8 &
\cmark & 0 & 0 &
\cmark & 0 & 0 &
\cmark & 9e-5 & 5e-8 &
20 & 20 & 0 & 15.0 \\

\texttt{ case3375wp\_k} & 29 & 0.0015 &
2 & 3e-1 & 5e-4 &
\cmark & 3e-2 & 7e-5 &
\cmark & 1e-8 & 2e-10 &
\cmark & 2e-7 & 1e-8 &
\cmark & 4e-7 & 5e-9 &
\cmark & 0 & 0 &
2 & 8e-2 & 3e-5 &
12 & 11 & 1 & 14.5 \\

\midrule
\rowcolor{largescale}
\multicolumn{28}{c}{\textbf{Large--Scale \texttt{pglib\_opf} Systems (4601--10480 buses)}} \\
\midrule

\texttt{ case4601\_goc} & 48 & 0.0004 &
4 & 4e-1 & 3e-4 &
\cmark & 6e-2 & 5e-5 &
\cmark & 1e-8 & 1e-10 &
\cmark & 1e-7 & 1e-8 &
\cmark & 1e-8 & 1e-9 &
\cmark & 0 & 0 &
\cmark & 0 & 0 &
1 & 1 & 0 & 9.6 \\

\texttt{ case5658\_epigrids} & 7 & 0.0001 &
\cmark & 8e-2 & 7e-5 &
\cmark & 8e-3 & 7e-6 &
\cmark & 1e-8 & 6e-11 &
\cmark & 1e-7 & 1e-8 &
\cmark & 1e-8 & 2e-9 &
\cmark & 0 & 0 &
\cmark & 0 & 0 &
0 & 0 & 0 & 7.9 \\

\texttt{ case7336\_epigrids} & 57 & 0.0004 &
5 & 1e-1 & 3e-4 &
\cmark & 4e-2 & 4e-5 &
\cmark & 1e-8 & 1e-10 &
\cmark & 1e-7 & 1e-8 &
\cmark & 3e-8 & 3e-9 &
\cmark & 0 & 0 &
\cmark & 0 & 0 &
2 & 2 & 0 & 23.9 \\

\texttt{ case8387\_pegase} & 61 & 0.0026 &
1 & 1e-1 & 1e-4 &
\cmark & 2e-2 & 2e-5 &
\cmark & 6e-9 & 6e-12 &
\cmark & 1e-6 & 3e-9 &
\cmark & 0 & 0 &
\cmark & 0 & 0 &
1 & 3e-2 & 2e-6 &
690 & 691 & 1 & 30.0 \\

\texttt{ case9591\_goc} & 23 & 0.0000 &
\cmark & 3e-4 & 3e-7 &
\cmark & 2e-5 & 3e-8 &
\cmark & 1e-8 & 2e-11 &
\cmark & 1e-7 & 2e-8 &
\cmark & 4e-8 & 3e-9 &
\cmark & 0 & 0 &
\cmark & 0 & 0 &
0 & 0 & 0 & 19.3 \\

\texttt{ case10480\_goc} & 58 & 0.0000 &
\cmark & 4e-4 & 6e-7 &
\cmark & 1e-4 & 3e-7 &
\cmark & 1e-8 & 3e-11 &
\cmark & 1e-7 & 2e-8 &
\cmark & 3e-8 & 3e-9 &
\cmark & 0 & 0 &
\cmark & 6e-5 & 5e-9 &
7 & 7 & 0 & 25.7 \\

\bottomrule
\end{tabular}%
}
\label{tab:allpass_global_summary}
\end{table*}

\begin{table*}[!t]
\footnotesize
\setlength{\tabcolsep}{2.0pt}
\renewcommand{\arraystretch}{1.08}
\captionsetup{font={footnotesize}}
\caption{Variable mismatch between AC OPF and APF OPF across medium- and large-scale \texttt{pglib\_opf} systems}
\centering

\resizebox{\textwidth}{!}{%
\begin{tabular}{l|
cc cc cc cc cc cc|
cc cc cc cc cc cc}
\toprule

\rowcolor{headergray}
 & \multicolumn{24}{c}{Mismatch (pu / rad)} \\
\cmidrule(lr){2-25}

\rowcolor{headergray}
 & \multicolumn{12}{c|}{Medium--Scale Systems} &
   \multicolumn{12}{c}{Large--Scale Systems} \\
\cmidrule(lr){2-13}\cmidrule(lr){14-25}

\rowcolor{headergray}
 & \multicolumn{2}{c}{\texttt{case588}}
 & \multicolumn{2}{c}{\texttt{case1354}}
 & \multicolumn{2}{c}{\texttt{case2383}}
 & \multicolumn{2}{c}{\texttt{case2746}}
 & \multicolumn{2}{c}{\texttt{case2869}}
 & \multicolumn{2}{c|}{\texttt{case3375}}
 & \multicolumn{2}{c}{\texttt{case4601}}
 & \multicolumn{2}{c}{\texttt{case5658}}
 & \multicolumn{2}{c}{\texttt{case7336}}
 & \multicolumn{2}{c}{\texttt{case8387}}
 & \multicolumn{2}{c}{\texttt{case9591}}
 & \multicolumn{2}{c}{\texttt{case10480}} \\
\cmidrule(lr){2-3}\cmidrule(lr){4-5}\cmidrule(lr){6-7}\cmidrule(lr){8-9}
\cmidrule(lr){10-11}\cmidrule(lr){12-13}
\cmidrule(lr){14-15}\cmidrule(lr){16-17}\cmidrule(lr){18-19}
\cmidrule(lr){20-21}\cmidrule(lr){22-23}\cmidrule(lr){24-25}

\rowcolor{headergray}
 & Max & Avg & Max & Avg & Max & Avg & Max & Avg & Max & Avg & Max & Avg
 & Max & Avg & Max & Avg & Max & Avg & Max & Avg & Max & Avg & Max & Avg \\
\midrule

$p^g$ &
3e-2 & 1e-3 & 6e-3 & 5e-5 & 2e-3 & 2e-5 & 2e-5 & 4e-8 & 2e-3 & 2e-5 & 1e-1 & 1e-3
& 4e-3 & 3e-5 & 6e-4 & 1e-6 & 2e-2 & 1e-4 & 1e-0 & 2e-3 & 4e-6 & 2e-8 & 1e-3 & 2e-5 \\

$q^g$ &
8e-2 & 5e-3 & 4e-2 & 8e-4 & 9e-3 & 2e-4 & 6e-3 & 7e-5 & 4e-3 & 9e-5 & 4e-2 & 1e-3
& 2e-1 & 4e-3 & 3e-2 & 1e-3 & 1e-1 & 2e-3 & 2e-0 & 9e-3 & 4e-2 & 1e-3 & 1e-0 & 7e-3 \\

$V$ &
1e-3 & 6e-4 & 4e-4 & 1e-5 & 1e-4 & 3e-5 & 2e-6 & 2e-7 & 3e-5 & 1e-6 & 5e-4 & 6e-5
& 1e-3 & 1e-4 & 5e-5 & 2e-6 & 2e-4 & 9e-6 & 1e-2 & 8e-5 & 2e-6 & 4e-8 & 8e-5 & 6e-6 \\

$\theta$ &
6e-3 & 2e-3 & 2e-4 & 4e-5 & 3e-4 & 2e-4 & 4e-3 & 2e-3 & 1e-3 & 7e-5 & 8e-3 & 3e-3
& 3e-3 & 3e-3 & 1e-3 & 7e-4 & 4e-3 & 2e-3 & 2e-2 & 4e-4 & 5e-6 & 4e-6 & 1e-4 & 6e-6 \\

$P_{ij}/P_{ji}$ &
4e-2 & 2e-3 & 5e-3 & 1e-4 & 9e-3 & 6e-5 & 2e-1 & 2e-3 & 4e-2 & 1e-4 & 1e-1 & 2e-3
& 3e-1 & 6e-4 & 4e-2 & 2e-4 & 9e-2 & 7e-4 & 6e-1 & 7e-4 & 1e-4 & 4e-7 & 1e-2 & 9e-6 \\

$Q_{ij}/Q_{ji}$ &
7e-2 & 2e-3 & 1e-2 & 2e-4 & 8e-3 & 7e-5 & 2e-2 & 4e-4 & 4e-3 & 3e-5 & 3e-2 & 6e-4
& 2e-1 & 5e-4 & 4e-3 & 4e-5 & 2e-2 & 2e-4 & 7e-1 & 1e-3 & 6e-5 & 1e-7 & 4e-1 & 9e-5 \\

\bottomrule
\end{tabular}%
}

\label{tab:mismatch_summary_med_large}
\end{table*}

\begin{table*}[!t]
\footnotesize
\color{black}
\setlength{\tabcolsep}{2pt}
\renewcommand{\arraystretch}{1.08}
\captionsetup{font={footnotesize}}
\caption{\textcolor{black}{Iteration-count and time reductions
(AC OPF - APF OPF) showing improvements
in both metrics}}
\label{tab:iteration_time_reductions}
\centering

\resizebox{\textwidth}{!}{%
\begin{tabular}{l|*{13}{c}}
\toprule
\rowcolor{headergray}
Quantity
& \texttt{case5}
& \texttt{case9}
& \texttt{case588}
& \texttt{case1354}
& \texttt{case2383}
& \texttt{case2746}
& \texttt{case2869}
& \texttt{case3375}
& \texttt{case4601}
& \texttt{case5658}
& \texttt{case7336}
& \texttt{case8387}
& \texttt{case10480} \\
\midrule
Iteration reduction
& 2 & 3 & 850 & 531 & 761 & 511 & 257
& 689 & 508 & 13 & 746 & 3210 & 2557 \\
Time reduction (s)
& 0.039 & 0.081 & 24.995 & 41.427 & 112.284 & 88.662 & 51.528
& 139.071 & 184.108 & 4.275 & 421.442 & 2289.104 & 2850.151 \\
\bottomrule
\end{tabular}%
}

\end{table*}

\begin{table*}[!t]
\centering
\footnotesize
\color{black}
\setlength{\tabcolsep}{2pt}
\renewcommand{\arraystretch}{1.12}
\captionsetup{font={footnotesize},labelfont={color=black},
              textfont={color=black}}
\caption{Comparison of AC OPF and APF OPF with DC-PF pre-rotation
under varying loading conditions.}
\label{tab:load_variation}

\resizebox{\textwidth}{!}{%
\begin{tabular}{l|cccc|cccc|cccc|cccc|cccc|cccc|cccc}
\toprule
\rowcolor{headergray}
& \multicolumn{4}{c|}{70\% loading}
& \multicolumn{4}{c|}{80\% loading}
& \multicolumn{4}{c|}{90\% loading}
& \multicolumn{4}{c|}{100\% loading}
& \multicolumn{4}{c|}{110\% loading}
& \multicolumn{4}{c|}{120\% loading}
& \multicolumn{4}{c}{130\% loading} \\
\rowcolor{headergray}
Case
& $R_t$ & $d_f$ & AC & APF
& $R_t$ & $d_f$ & AC & APF
& $R_t$ & $d_f$ & AC & APF
& $R_t$ & $d_f$ & AC & APF
& $R_t$ & $d_f$ & AC & APF
& $R_t$ & $d_f$ & AC & APF
& $R_t$ & $d_f$ & AC & APF \\
\midrule
\texttt{case5}
& 1.35 & .006 & S & S
& 1.29 & .002 & S & S
& 1.32 & .000 & S & S
& 1.26 & .000 & S & S
& 1.33 & .000 & S & S
& 1.25 & .000 & S & S
& 1.17 & .000 & S & S \\
\texttt{case9}
& 1.62 & .000 & S & S
& 1.72 & .000 & S & S
& 1.86 & .000 & S & S
& 1.74 & .000 & S & S
& 1.45 & .000 & S & S
& 1.45 & .000 & S & S
& 1.53 & .000 & S & S \\
\texttt{case12da}
& 1.31 & .000 & S & S
& 1.22 & .000 & S & S
& 1.07 & .000 & S & S
& 1.10 & .000 & S & S
& 1.10 & .000 & S & S
& 1.10 & .000 & S & S
& 1.10 & .000 & S & S \\
\texttt{case18}
& 1.19 & .000 & S & S
& 1.14 & .000 & S & S
& 1.13 & .000 & S & S
& 1.08 & .000 & S & S
& 1.19 & .000 & S & S
& 1.17 & .000 & S & S
& 1.15 & .000 & S & S \\
\texttt{case22}
& 1.14 & .000 & S & S
& 1.14 & .000 & S & S
& 1.32 & .000 & S & S
& 1.18 & .000 & S & S
& 1.23 & .000 & S & S
& 1.22 & .000 & S & S
& 1.21 & .000 & S & S \\
\texttt{case24\_ieee}
& 2.05 & .000 & S & S
& 1.75 & .000 & S & S
& 1.48 & .000 & S & S
& 1.66 & .000 & S & S
& 1.44 & .000 & S & S
& -- & -- & I & I
& -- & -- & I & I \\
\texttt{case30}
& 1.39 & .000 & S & S
& 1.90 & .000 & S & S
& 1.80 & .000 & S & S
& 1.13 & .000 & S & S
& -- & -- & I & I
& -- & -- & I & I
& -- & -- & I & I \\
\texttt{case33bw}
& 1.29 & .000 & S & S
& 1.31 & .000 & S & S
& 1.24 & .000 & S & S
& 1.25 & .000 & S & S
& 1.23 & .000 & S & S
& -- & -- & I & I
& -- & -- & I & I \\
\texttt{case38si}
& 1.52 & .000 & S & S
& 1.33 & .000 & S & S
& 1.30 & .000 & S & S
& 1.22 & .000 & S & S
& 1.29 & .000 & S & S
& -- & -- & I & I
& -- & -- & I & I \\
\texttt{case39}
& 2.93 & .000 & S & S
& 2.10 & .000 & S & S
& 2.92 & .000 & S & S
& 1.88 & .000 & S & S
& -- & -- & I & I
& -- & -- & I & I
& -- & -- & I & I \\
\texttt{case51ga}
& 1.36 & .000 & S & S
& 1.33 & .000 & S & S
& 1.33 & .000 & S & S
& 1.44 & .000 & S & S
& -- & -- & I & I
& -- & -- & I & I
& -- & -- & I & I \\
\texttt{case57}
& 2.99 & .000 & S & S
& 3.39 & .000 & S & S
& 3.57 & .000 & S & S
& 3.67 & .000 & S & S
& -- & -- & I & I
& -- & -- & I & I
& -- & -- & I & I \\
\texttt{case69}
& 1.35 & .000 & S & S
& 1.40 & .000 & S & S
& 1.34 & .000 & S & S
& 1.37 & .000 & S & S
& -- & -- & S & I
& -- & -- & I & I
& -- & -- & I & I \\
\texttt{case74ds}
& 1.37 & .000 & S & S
& 1.36 & .000 & S & S
& 1.36 & .000 & S & S
& 1.34 & .000 & S & S
& 1.39 & .000 & S & S
& 1.38 & .000 & S & S
& 1.34 & .000 & S & S \\
\texttt{case118}
& 1.36 & .000 & S & S
& -- & -- & S & F
& 2.11 & .000 & S & S
& -- & -- & F & S
& 1.08 & .000 & S & S
& -- & -- & S & F
& -- & -- & F & S \\
\texttt{case141}
& 1.42 & .000 & S & S
& 1.32 & .000 & S & S
& 1.34 & .000 & S & S
& 1.33 & .000 & S & S
& 1.36 & .000 & S & S
& 1.29 & .000 & S & S
& 1.31 & .000 & S & S \\
\texttt{case300}
& -- & -- & S & I
& 1.75 & .139 & S & S
& -- & -- & F & S
& 1.61 & .000 & S & S
& -- & -- & I & I
& -- & -- & F & I
& -- & -- & F & I \\
\midrule
\texttt{case588}
& -- & -- & F & S
& -- & -- & F & S
& 1.93 & .004 & S & S
& 1.26 & .002 & S & S
& .34 & .000 & S & S
& -- & -- & F & S
& -- & -- & F & S \\
\texttt{case1354}
& 4.71 & .014 & S & S
& 1.46 & .006 & S & S
& 2.15 & .002 & S & S
& 1.40 & .000 & S & S
& .73 & .000 & S & S
& -- & -- & I & I
& -- & -- & I & I \\
\texttt{case2383}
& 2.30 & .000 & S & S
& .46 & .000 & S & S
& .35 & .000 & S & S
& 2.04 & .000 & S & S
& -- & -- & I & I
& -- & -- & F & F
& -- & -- & F & F \\
\texttt{case2746}
& -- & -- & I & I
& .75 & .000 & S & S
& 1.89 & .000 & S & S
& 1.42 & .000 & S & S
& -- & -- & I & I
& -- & -- & F & I
& -- & -- & F & F \\
\texttt{case2869}
& 1.35 & .012 & S & S
& .94 & .003 & S & S
& -- & -- & S & F
& 1.68 & .000 & S & S
& -- & -- & F & F
& -- & -- & F & F
& -- & -- & F & F \\
\texttt{case3375}
& -- & -- & I & F
& -- & -- & I & I
& -- & -- & I & I
& 1.38 & .002 & S & S
& -- & -- & F & F
& -- & -- & I & I
& -- & -- & F & F \\
\texttt{case4601}
& .77 & .000 & S & S
& 1.60 & .000 & S & S
& .48 & .000 & S & S
& 1.87 & .000 & S & S
& -- & -- & I & I
& -- & -- & I & I
& -- & -- & I & F \\
\texttt{case5658}
& 1.07 & .006 & S & S
& .71 & .003 & S & S
& .96 & .001 & S & S
& 1.09 & .000 & S & S
& .79 & .000 & S & S
& -- & -- & I & I
& -- & -- & I & I \\
\texttt{case7336}
& .80 & .002 & S & S
& 1.23 & .002 & S & S
& 1.40 & .001 & S & S
& 2.88 & .000 & S & S
& .48 & .000 & S & S
& -- & -- & I & F
& -- & -- & F & F \\
\texttt{case8387}
& -- & -- & F & F
& -- & -- & F & F
& -- & -- & F & F
& -- & -- & F & F
& -- & -- & I & I
& -- & -- & I & I
& -- & -- & F & F \\
\texttt{case9591}
& 1.31 & .000 & S & S
& 1.08 & .001 & S & S
& 1.26 & .046 & S & S
& .98 & .740 & S & S
& -- & -- & I & F
& -- & -- & F & F
& -- & -- & I & F \\
\bottomrule
\end{tabular}%
}

\vspace{2pt}
\begin{minipage}{\textwidth}
\scriptsize
\textbf{Metrics:}
$R_t=t_{\mathrm{AC}}/t_{\mathrm{APF}}$ is the solver-time ratio;
$R_t>1$ indicates faster APF OPF.
$d_f=100|f_{\mathrm{APF}}-f_{\mathrm{AC}}|/|f_{\mathrm{AC}}|$
is the absolute relative objective difference (\%).

\textbf{Termination status:}
AC and APF columns report the respective solver statuses:
S = success, including acceptable-level termination;
I = reported infeasibility;
F = other unsuccessful termination.
Numerical comparisons are shown only when both runs report S.
Values of .000 are rounded and do not imply exact equality.

\end{minipage}
\end{table*}

\begin{table*}[!t]
\centering
\footnotesize
\color{black}
\setlength{\tabcolsep}{1.5pt}
\renewcommand{\arraystretch}{1.15}
\captionsetup{font={footnotesize},labelfont={color=black},textfont={color=black}}
\caption{Sensitivity of APF OPF with DC-PF pre-rotation to parameter $a$.}
\label{tab:a_sensitivity}
\resizebox{\textwidth}{!}{%
\begin{tabular}{c|rrr|rrr|rrr|rrr|rrr|rrr|rrr|rrr|rrr|rrr|rrr|rrr}
\toprule
\rowcolor{headergray}
$a$ & \multicolumn{3}{c|}{\texttt{case9}} & \multicolumn{3}{c|}{\texttt{case24}} & \multicolumn{3}{c|}{\texttt{case57}} & \multicolumn{3}{c|}{\texttt{case118}} & \multicolumn{3}{c|}{\texttt{case300}} & \multicolumn{3}{c|}{\texttt{case588}} & \multicolumn{3}{c|}{\texttt{case1354}} & \multicolumn{3}{c|}{\texttt{case2383wp}} & \multicolumn{3}{c|}{\texttt{case2746wp}} & \multicolumn{3}{c|}{\texttt{case2869}} & \multicolumn{3}{c|}{\texttt{case3375wp}} & \multicolumn{3}{c}{\texttt{case4601}} \\
\rowcolor{headergray}
 & $t$ & $d_f$ & $M$ & $t$ & $d_f$ & $M$ & $t$ & $d_f$ & $M$ & $t$ & $d_f$ & $M$ & $t$ & $d_f$ & $M$ & $t$ & $d_f$ & $M$ & $t$ & $d_f$ & $M$ & $t$ & $d_f$ & $M$ & $t$ & $d_f$ & $M$ & $t$ & $d_f$ & $M$ & $t$ & $d_f$ & $M$ & $t$ & $d_f$ & $M$ \\
\midrule
0.05 & 0.02 & 0.0 & 5.6 & 0.05 & 0.0 & 9.3 & 0.09 & 0.0 & 24.4 & 1.21 & 0.0 & 14.5 & 2.23 & 0.0 & 16.5 & 448.15 & 31.1 & 132.0 & 45.05 & 3.7 & 231.0 & 134.00 & 15.4 & 202.0 & 131.62 & 0.0 & 107.0 & 128.84 & 1.5 & 259.0 & 2574.76 & 12.2 & 166.0 & 2019.88 & 59.0 & 311.0 \\
0.10 & 0.02 & 0.0 & 2.5 & 0.06 & 0.0 & 4.1 & 0.09 & 0.0 & 11.7 & 1.30 & 0.0 & 6.6 & 1.35 & 0.0 & 7.3 & 83.01 & 3.6 & 146.0 & 241.73 & 0.0 & 167.0 & 670.98 & 0.1 & 133.0 & 256.42 & 0.0 & 48.1 & 160.41 & 0.1 & 144.0 & 573.21 & 7.0 & 103.0 & 1398.72 & 69.5 & 183.0 \\
0.15 & 0.02 & 0.0 & 1.4 & 0.07 & 0.0 & 2.4 & 0.08 & 0.0 & 6.9 & 1.36 & 0.0 & 3.8 & 1.36 & 0.0 & 4.2 & 161.73 & 1.1 & 87.3 & 54.64 & 0.0 & 98.1 & 249.50 & 0.0 & 84.3 & 121.86 & 0.0 & 28.1 & 113.08 & 0.0 & 84.4 & 469.08 & 3.8 & 68.2 & 357.87 & 76.6 & 110.0 \\
0.20 & 0.02 & 0.0 & 0.9 & 0.07 & 0.0 & 1.5 & 0.10 & 0.0 & 4.5 & 2.64 & 0.0 & 2.5 & 1.63 & 0.0 & 2.7 & 124.04 & 0.1 & 54.8 & 312.82 & 0.0 & 63.2 & 207.52 & 0.0 & 54.6 & 255.29 & 0.0 & 18.1 & 154.43 & 0.0 & 54.4 & 532.73 & 0.5 & 45.7 & 449.73 & 2.8 & 88.5 \\
0.25 & 0.02 & 0.0 & 0.6 & 0.08 & 0.0 & 1.0 & 0.09 & 0.0 & 3.0 & 3.18 & 0.0 & 1.6 & 2.82 & 0.0 & 1.8 & 147.72 & 0.4 & 36.3 & 78.33 & 0.0 & 42.2 & 232.76 & 0.0 & 36.5 & 214.05 & 0.0 & 12.0 & 129.35 & 0.0 & 36.2 & 383.23 & 0.0 & 33.7 & 252.46 & 0.0 & 78.3 \\
0.30 & 0.02 & 0.0 & 0.4 & 0.06 & 0.0 & 0.7 & 0.10 & 0.0 & 2.0 & 2.71 & 0.0 & 1.1 & 3.07 & 0.0 & 1.2 & 173.44 & 0.0 & 24.3 & 44.77 & 0.0 & 28.1 & 191.29 & 0.0 & 24.4 & 181.07 & 0.0 & 8.0 & 121.31 & 0.0 & 24.2 & 238.79 & 0.0 & 22.7 & 454.24 & 0.0 & 52.5 \\
0.35 & 0.02 & 0.0 & 0.3 & 0.07 & 0.0 & 0.4 & 0.10 & 0.0 & 1.3 & 2.20 & 0.0 & 0.7 & 1.71 & 0.0 & 0.8 & 110.06 & 0.0 & 15.6 & 58.15 & 0.0 & 18.1 & 176.72 & 0.0 & 15.7 & 208.34 & 0.0 & 5.2 & 156.34 & 0.0 & 15.5 & 376.81 & 0.0 & 14.8 & 203.11 & 0.0 & 33.9 \\
0.40 & 0.02 & 0.0 & 0.2 & 0.07 & 0.0 & 0.3 & 0.10 & 0.0 & 0.8 & 1.97 & 0.0 & 0.4 & 3.29 & 0.0 & 0.4 & 90.13 & 0.0 & 9.1 & 149.62 & 0.0 & 10.6 & 191.82 & 0.0 & 9.2 & 142.02 & 0.0 & 3.0 & 128.75 & 0.0 & 9.1 & 1038.39 & 0.0 & 8.7 & 250.88 & 0.0 & 19.9 \\
0.45 & 0.02 & 0.0 & 0.1 & 0.08 & 0.0 & 0.1 & 0.10 & 0.0 & 0.3 & 1.27 & 0.0 & 0.2 & 2.52 & 0.0 & 0.2 & 110.93 & 0.0 & 4.1 & 72.41 & 0.0 & 4.7 & 202.73 & 0.0 & 4.1 & 207.65 & 0.0 & 1.3 & 116.38 & 0.0 & 4.0 & 465.90 & 0.0 & 4.0 & 174.04 & 0.0 & 8.9 \\
\rowcolor{headergray}
\textbf{0.50} & 0.03 & 0.0 & $5.4\mathrm{e}{-5}$ & 0.06 & 0.0 & $5.5\mathrm{e}{-5}$ & 0.06 & 0.0 & 0.0 & 0.92 & 0.0 & 0.0 & 3.47 & 0.0 & 0.0 & 121.60 & 0.0 & 0.0 & 128.64 & 0.0 & 0.0 & 171.57 & 0.0 & 0.0 & 152.20 & 0.0 & 0.0 & 103.92 & 0.0 & 0.0 & 747.48 & 0.0 & 0.3 & 237.66 & 0.0 & 0.4 \\
0.55 & 0.03 & 0.0 & 0.1 & 0.07 & 0.0 & 0.1 & 0.09 & 0.0 & 0.3 & 1.96 & 0.0 & 0.2 & 2.95 & 0.0 & 0.2 & 75.40 & 0.0 & 3.3 & 115.24 & 0.0 & 3.8 & 247.26 & 0.0 & 3.3 & 127.12 & 0.0 & 1.1 & 108.67 & 0.0 & 3.3 & 242.53 & 0.0 & 2.8 & 207.14 & 0.0 & 7.0 \\
0.60 & 0.03 & 0.0 & 0.1 & 0.06 & 0.0 & 0.2 & 0.08 & 0.0 & 0.5 & 1.80 & 0.0 & 0.3 & 2.98 & 0.0 & 0.3 & 76.40 & 0.0 & 6.0 & 77.85 & 0.0 & 7.0 & 137.66 & 0.0 & 6.0 & 138.10 & 0.0 & 2.0 & 111.02 & 0.0 & 6.0 & 899.06 & 0.0 & 5.4 & 247.34 & 0.0 & 13.0 \\
0.65 & 0.02 & 0.0 & 0.1 & 0.05 & 0.0 & 0.2 & 0.08 & 0.0 & 0.7 & 2.21 & 0.0 & 0.4 & 2.58 & 0.0 & 0.4 & 50.57 & 0.0 & 8.4 & 102.43 & 0.0 & 9.7 & 228.91 & 0.0 & 8.4 & 245.06 & 0.0 & 2.8 & 117.59 & 0.0 & 8.4 & 463.92 & 0.0 & 7.5 & 251.98 & 0.0 & 18.1 \\
0.70 & 0.03 & 0.0 & 0.2 & 0.06 & 0.0 & 0.3 & 0.08 & 0.0 & 0.9 & 2.49 & 0.0 & 0.5 & 1.77 & 0.0 & 0.5 & 41.91 & 0.0 & 10.4 & 100.78 & 0.0 & 12.0 & 151.73 & 0.0 & 10.4 & 156.16 & 0.0 & 3.4 & 249.00 & 0.0 & 10.3 & 251.52 & 0.0 & 9.4 & 571.49 & 0.0 & 22.5 \\
0.75 & 0.02 & 0.0 & 0.2 & 0.08 & 0.0 & 0.3 & 0.08 & 0.0 & 1.0 & 1.90 & 0.0 & 0.6 & 1.41 & 0.0 & 0.6 & 31.54 & 0.0 & 12.1 & 80.18 & 0.0 & 14.0 & 238.01 & 0.0 & 12.1 & 105.51 & 0.0 & 4.0 & 113.95 & 0.0 & 12.1 & 111.60 & 0.0 & 11.0 & 203.94 & 0.0 & 26.3 \\
0.80 & 0.02 & 0.0 & 0.2 & 0.08 & 0.0 & 0.4 & 0.10 & 0.0 & 1.1 & 2.08 & 0.0 & 0.6 & 3.66 & 0.0 & 0.7 & 28.42 & 0.0 & 13.6 & 75.46 & 0.0 & 15.8 & 252.41 & 0.0 & 13.6 & 159.93 & 0.0 & 4.5 & 110.19 & 0.0 & 13.6 & 166.38 & 0.0 & 12.4 & 142.09 & 0.0 & 29.6 \\
0.85 & 0.02 & 0.0 & 0.3 & 0.08 & 0.0 & 0.4 & 0.09 & 0.0 & 1.2 & 3.51 & 0.0 & 0.7 & 2.64 & 0.0 & 0.7 & 50.37 & 0.0 & 14.9 & 75.59 & 0.0 & 17.3 & 186.10 & 0.0 & 15.0 & 149.76 & 0.0 & 5.0 & 115.38 & 0.0 & 14.9 & 414.63 & 0.0 & 13.6 & 223.27 & 0.0 & 32.5 \\
0.90 & 0.02 & 0.0 & 0.3 & 0.08 & 0.0 & 0.5 & 0.11 & 0.0 & 1.4 & 1.70 & 0.0 & 0.7 & 2.46 & 0.0 & 0.8 & 97.12 & 0.0 & 16.1 & 92.31 & 0.0 & 18.7 & 208.38 & 0.0 & 16.2 & 124.42 & 0.0 & 5.4 & 137.40 & 0.0 & 16.1 & 224.69 & 0.0 & 14.7 & 248.83 & 0.0 & 35.1 \\
0.95 & 0.02 & 0.0 & 0.3 & 0.06 & 0.0 & 0.5 & 0.08 & 0.0 & 1.4 & 1.81 & 0.0 & 0.8 & 3.01 & 0.0 & 0.8 & 42.73 & 0.0 & 17.2 & 144.00 & 0.0 & 19.9 & 206.32 & 0.0 & 17.2 & 124.35 & 0.0 & 5.7 & 110.88 & 0.0 & 17.1 & 609.58 & 0.0 & 15.7 & 278.69 & 0.0 & 37.4 \\
\bottomrule
\end{tabular}%
}
\vspace{2pt}
\begin{minipage}{\textwidth}
\scriptsize
$t$: APF solver-call time (s), excluding model construction and DC-reference computation;
$M$: max mismatch (pu), defined as $\max\{\max_i|P_i^{\mathrm{mis}}|,\max_i|Q_i^{\mathrm{mis}}|\}$
using the exact AC equations at the returned APF operating point.
All runs are included.
\end{minipage}
\end{table*}

\subsubsection{Small Systems (9--300 Buses)}

This category includes standard IEEE and MATPOWER test cases ranging from \texttt{case9} to \texttt{case300}. These networks are relatively small and usually have mild nonlinear behavior, limited congestion, and well-conditioned operating points. All experiments for small-scale systems are performed using YALMIP modeling.

\paragraph{Speedup}
For all small-scale systems, the APF formulation shows a clear reduction in computational time compared to the classical AC OPF. According to Table~\ref{tab:allpass_global_summary}, the time speedup varies from about $11\%$ in \texttt{case30} to more than $70\%$ in \texttt{case57}, with most cases achieving speedups between $20\%$ and $45\%$. These improvements are obtained without changing solver tolerances or stopping criteria. Therefore, the speedup mainly comes from better numerical conditioning of the problem introduced by the APF kernel, rather than changes in solver settings.

\paragraph{Optimality Gap}
For all small-scale test cases, the APF formulation produces objective values that are identical to those of the classical AC OPF within numerical precision. As shown in Table~\ref{tab:allpass_global_summary}, the optimality gap is zero for all systems in this group. This indicates that the APF kernel does not affect the optimal cost or the optimal operating point of the problem.

\paragraph{Feasibility}
All APF solutions for small networks remain feasible when they are evaluated using the exact nonlinear AC power flow equations and constraints. The active and reactive power balance mismatches at buses are very small and always below the specified tolerances. Voltage magnitude limits, generator active and reactive power limits, branch thermal limits, and angle difference constraints are all satisfied, with no violations reported in Table~\ref{tab:allpass_global_summary}. These results confirm that the APF solutions are physically consistent with the original AC OPF model for all small-scale cases.

\paragraph{Congestion}
Most small-scale networks do not have congested transmission lines. In cases where congestion exists, such as \texttt{case30}, the APF formulation identifies exactly the same congested branches as the AC OPF solution. As shown by the zero congestion mismatch values in Table~\ref{tab:allpass_global_summary}, there is no difference between the two formulations in terms of congestion.

\paragraph{Max $\theta_{i}-\theta_{j}$}
The maximum branch-wise voltage angle differences remain within reasonable physical limits for all small-scale systems. For distribution-type networks, these angle differences are typically below $1^\circ$, while for more meshed transmission networks they are larger. The highest observed value is about $20.7^\circ$ in \texttt{case300}. The close agreement between the AC OPF and APF OPF results shows that the APF kernel preserves the angle behavior of the original AC power flow model.

\subsubsection{Medium Systems (588--3375 Buses)}

The medium-scale category consists of more challenging PGLib transmission networks with denser meshed topology and stronger nonlinear coupling. This group includes \texttt{case588\_sdet}, \texttt{case1354\_pegase}, \texttt{case2383wp\_k}, \texttt{case2746wp\_k}, \texttt{case2869\_pegase}, and \texttt{case3375wp\_k}. All medium-scale experiments are conducted using the CasADi-based implementation.

\paragraph{Speedup}
For medium-scale systems, the APF formulation consistently achieves noticeable reductions in computational time compared to the classical AC OPF. As reported in Table~\ref{tab:allpass_global_summary}, time speedup values range from approximately $20\%$ to $50\%$, with most cases showing speedups between $27\%$ and $40\%$. The largest improvements are observed in cases such as \texttt{case2383wp\_k} and \texttt{case2869\_pegase}. These results indicate that the APF kernel improves numerical conditioning of the nonlinear equations, leading to faster convergence without modifying solver settings.

\paragraph{Optimality Gap}
Across all medium-scale networks, the optimality gap remains very small. As shown in Table~\ref{tab:allpass_global_summary}, gap values are either zero or below $2\times10^{-3}\%$. This confirms that the APF formulation preserves the objective value of the AC OPF even for larger and more nonlinear systems, and that the economic dispatch structure remains essentially unchanged.

\paragraph{Feasibility}
Most medium-scale APF solutions remain feasible when evaluated under the exact nonlinear AC power flow equations. Bus active and reactive power balance mismatches are generally below the prescribed tolerances. In a few stressed cases, such as \texttt{case2746wp\_k} and \texttt{case3375wp\_k}, small active-power balance violations are observed at a limited number of buses. These violations are localized and relatively small in magnitude, while voltage magnitudes, generator limits, thermal limits, and angle-difference constraints remain satisfied. Overall, the feasibility results indicate strong consistency between the APF and AC formulations.

\paragraph{Congestion}
For most medium-scale networks, the APF formulation reproduces the same congestion pattern as the classical AC OPF. Exact congestion matching is observed in all cases except \texttt{case3375wp\_k}, where a single mismatched congested branch is reported. Despite this minor discrepancy, the overall congestion structure and stress locations in the network remain highly consistent between the two formulations.

\paragraph{Max $\theta_{i}-\theta_{j}$}
Maximum branch-wise voltage angle differences for medium-scale systems typically range between $11^\circ$ and $15^\circ$, as reported in Table~\ref{tab:allpass_global_summary}. These values are significantly larger than those in small-scale networks and provide a meaningful test of the APF approximation beyond small-angle regimes. The close agreement in angle behavior, together with the small variable mismatches reported in Table~\ref{tab:mismatch_summary_med_large}, demonstrates that the APF kernel accurately captures nonlinear angle interactions in meshed transmission networks.

\paragraph{Variable Mismatch Analysis}
For all cases, mismatches in generator dispatch, voltage magnitudes, voltage angles, and branch power flows remain small. Average mismatches are typically several orders of magnitude lower than the corresponding maximum values, indicating that deviations are limited and localized. Even in more stressed networks, the APF formulation closely tracks the AC OPF solution, confirming strong variable-level agreement between the two models.

\subsubsection{Large Systems (4601--10{,}480 Buses)}

The large-scale category contains the most challenging benchmark systems, including \texttt{case4601\_goc}, \texttt{case5658\_epigrids}, \texttt{case7336\_epigrids}, \texttt{case8387\_pegase}, \texttt{case9591\_goc}, and \texttt{case10480\_goc}. These networks feature large sizes, dense interconnections, and stronger nonlinear effects. All large-scale experiments are conducted using the CasADi-based implementation.

\paragraph{Speedup}
For large-scale systems, the APF formulation continues to provide noticeable computational benefits compared to the classical AC OPF. As shown in Table~\ref{tab:allpass_global_summary}, time speedup values range from relatively modest improvements of about $7\%$ in \texttt{case5658\_epigrids} to more significant gains above $50\%$ in cases such as \texttt{case7336\_epigrids} and \texttt{case10480\_goc}. These results indicate that the APF kernel remains effective in improving numerical conditioning even at very large problem sizes.

\paragraph{Optimality Gap}
Across all large-scale cases, the optimality gap remains very small. For most systems, the gap is either zero or below $10^{-3}\%$, as reported in Table~\ref{tab:allpass_global_summary}. In particular, \texttt{case9591\_goc} and \texttt{case10480\_goc} exhibit zero optimality gap. This demonstrates that the APF formulation preserves objective accuracy even for networks with more than ten thousand buses.

\paragraph{Feasibility}
Several large-scale networks satisfy all true AC feasibility checks, including \texttt{case5658\_epigrids}, \texttt{case9591\_goc}, and \texttt{case10480\_goc}. In other cases, small active-power balance violations are observed at a limited number of buses, such as in \texttt{case4601\_goc} and \texttt{case7336\_epigrids}. \textcolor{black}{For \texttt{case8387\_pegase}, both AC OPF and APF OPF
encounter convergence difficulties. APF OPF terminates sooner with a final operating point close to that returned by AC OPF, although feasibility and congestion discrepancies remain.}

\paragraph{Congestion}
For most large-scale networks, the APF formulation reproduces the same congestion pattern as the AC OPF. Exact congestion matching is observed in fully feasible cases and in several stressed cases, including \texttt{case4601\_goc} and \texttt{case7336\_epigrids}. A small congestion mismatch is only reported for \texttt{case8387\_pegase}, where one congested line differs between the two formulations, while the overall congestion structure remains largely consistent.

\paragraph{Max $\theta_{i}-\theta_{j}$}
Maximum branch-wise voltage angle differences in large-scale systems range from approximately $8^\circ$ to $30^\circ$, with the largest value observed in \texttt{case8387\_pegase}. These large angle spreads provide a stringent test for the APF approximation well beyond small-angle operating regimes. The close agreement between the AC OPF and APF OPF results confirms that the APF kernel can accurately capture nonlinear angular behavior even in very large and heavily loaded networks.

\paragraph{Variable Mismatch Analysis}
Overall, mismatches in generator dispatch, voltage magnitudes, voltage angles, and branch power flows remain small. Average mismatches are consistently much smaller than the corresponding maximum values, indicating that deviations are limited and localized. Larger mismatches are mainly observed in the most stressed case, \texttt{case8387\_pegase}, which also exhibits feasibility sensitivity in the AC OPF solution. These results show that, even at extreme scales, the APF formulation closely follows the AC OPF solution at the variable level.

\subsubsection{Overall Discussion}

Overall, the APF formulation consistently improves computational performance across all network sizes while maintaining close agreement with the classical AC OPF in terms of objective value, feasibility under true AC evaluation, congestion patterns, and voltage angle behavior. Larger speedups are generally observed for more complex and larger networks, reflecting improved numerical conditioning. \textcolor{black}{For \texttt{case8387\_pegase}, the comparison concerns
the final operating points and computation times of two runs that encountered convergence difficulties, without identifying their cause.} \textcolor{black}{Table~\ref{tab:iteration_time_reductions} reports
the reductions in IPOPT iteration counts and solver times (AC OPF - APF OPF) showing improvements in both metrics.}

\textcolor{black}{To examine performance beyond the default loading conditions, Table~\ref{tab:load_variation} compares AC OPF and APF OPF with DC-PF pre-rotation across 28 systems at seven loading levels, from 70\% to 130\% of the original demand. Both active and reactive demands are scaled by the same factor, while generation, voltage, and branch limits remain unchanged. The results show that APF OPF achieves shorter solver times in many scenarios where both formulations terminate successfully, while generally maintaining small objective differences. For example, at 70\% loading, APF OPF is 4.71 times faster for \texttt{case1354}, with an objective difference of 0.014\%. However, the computational benefit depends on the system and loading level: for \texttt{case588} at 110\% loading, the solver-time ratio is 0.34, indicating a longer APF OPF solve despite closely matching objective values. The status columns also reveal differences in convergence behavior. APF OPF reports successful termination for \texttt{case588} at several loading levels where AC OPF terminates unsuccessfully, whereas the opposite occurs for other scenarios, such as \texttt{case300} at 70\% loading. At higher loading levels, many systems exhibit solver-reported infeasibility or other termination difficulties in both formulations; unsuccessful runs also occur at light loading in some systems. Numerical time and objective comparisons are therefore reported only when both runs terminate successfully. These results extend the evaluation to light and heavy loading across different network sizes and show both the computational benefits and the dependence of convergence behavior on operating conditions.}

\textcolor{black}{Table~\ref{tab:a_sensitivity} examines the sensitivity of APF OPF to the parameter $a$ across 12 benchmark systems ranging from 9 to 4,601 buses. For each system, the exact AC OPF formulation is solved once as a reference, and APF OPF is solved for 19 values of $a$, from 0.05 to 0.95 in increments of 0.05. All runs use CasADi and IPOPT, the same case-based initial operating point, and identical solver settings without CPU-time or wall-time limits. The table reports APF solver time, excluding model construction and DC-reference computation; the absolute relative objective difference from the AC reference; and the maximum active or reactive nodal power-balance mismatch evaluated using the exact AC equations. Objective differences below $10^{-3}\%$ are displayed as 0.0. Results from unsuccessful runs describe their final iterates and time to termination; the AC references for \texttt{case118} and \texttt{case300} terminated with step-computation errors. Computational performance varies with both the system and $a$, without a consistent monotonic relationship between parameter value and solver time. Small objective differences occur over a broad range of parameter values, but can coexist with substantial exact AC power-balance mismatches. In contrast, $a=0.5$ gives the smallest reported maximum mismatch in every tested system. This supports selecting $a=0.5$ for its stronger agreement with the exact AC equations. Together with the shorter solver times relative to classical AC OPF reported in Table~\ref{tab:allpass_global_summary}, these observations support this choice as a practical balance between accuracy and computational efficiency: it need not be the fastest among all tested values of $a$ to provide a computational benefit over the classical formulation.}

\subsection{Comparison with Cartesian AC OPF}
\label{subsec:cartesian}

{\color{black}
Cartesian coordinate AC OPF expresses voltages as $e_i+\mathbf{j}f_i$,
eliminating trigonometric functions from the power balance equations
at the cost of increased nonlinearity elsewhere. Voltage magnitude
bounds become quadratic inequalities, angle-difference bounds become
bilinear, thermal limits become degree-4 polynomials in $(e_i,f_i)$,
and trigonometric functions reappear in angle-difference constraints
when enforced exactly. In APF OPF, all constraints retain their
standard polar forms. The DC pre-rotation step, which centers the
rational kernel at the expected operating point, has no analog in
Cartesian coordinates.

\subsubsection*{Feasible Set Geometry}

The coordinate choice fundamentally determines the topology of the
feasible set and the behavior of interior-point
solvers~\cite{hiskens2001exploring}. In polar coordinates, voltage
and angle bounds are independent box constraints, so the feasible
set is a convex rectangle in $(V_2,\delta_2)$ space with no holes
and a unique local minimum, as shown in
Fig.~\ref{fig:feasible_geometry}(a) for a two-bus network with
$G=0.1$\,pu, $B=-1.0$\,pu, $[V_{\min},V_{\max}]=[0.9,1.1]$\,pu,
$|\delta|\le60^\circ$, and $\overline{S}=0.8$\,pu. In Cartesian
coordinates, the same voltage bounds define a circular annulus in
$(e_2,f_2)$ space, Fig.~\ref{fig:feasible_geometry}(b), which is
non-convex and not simply connected.
The cost function becomes quartic in $(e_i,f_i)$, and the topology
admits spurious KKT points that are not globally
optimal~\cite{lavaei2011zero}. APF OPF
preserves the convex polar geometry, and since the two formulations
optimize over topologically distinct feasible sets, differences in
convergence behavior reflect geometry as much as formulation
structure.

\begin{figure}[!t]
\centering
\captionsetup{font={footnotesize}}
\includegraphics[width=\columnwidth]{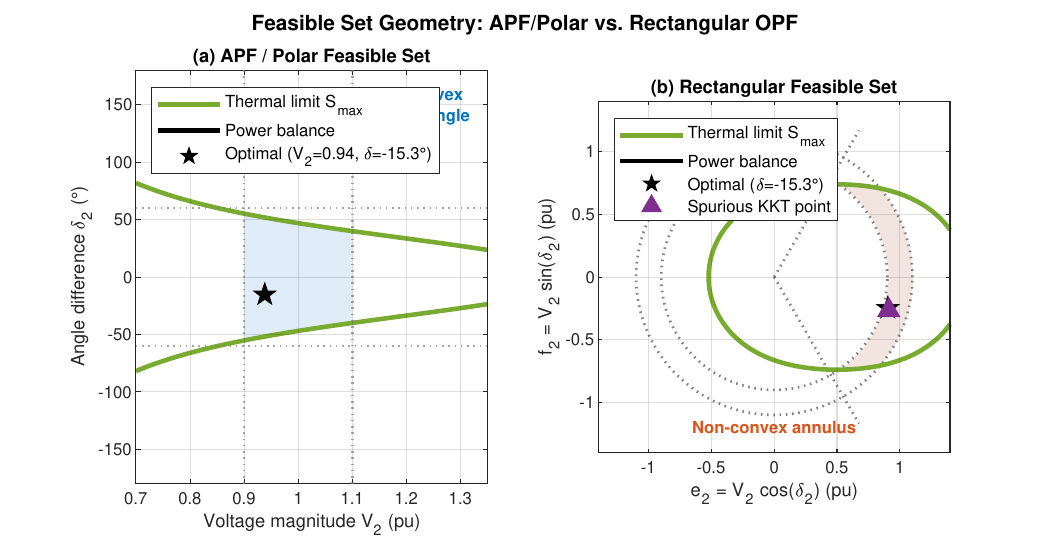}
\caption{{\color{black}Feasible set geometry}}
\label{fig:feasible_geometry}
\end{figure}

\subsubsection*{Barrier Function Structure}

In IPM, each inequality constraint contributes a barrier Hessian
term to the KKT system. For the
voltage lower bound in polar coordinates,
$g(V_i)=\underline{V}_i-V_i$, so $\|\nabla g\|=1$ identically and
the barrier is a scalar perturbation in the $V_i$ direction only
— fully decoupled from all other variables at every operating point.

In Cartesian coordinates, the same bound gives
$g(e_i,f_i)=\underline{V}_i^2-(e_i^2+f_i^2)$, so
$\nabla g=-2(e_i,f_i)^T$. The resulting barrier Hessian block
$4\mu V_i^2(e_i,f_i)(e_i,f_i)^T/s^2$ is rank-1, meaning one
direction in $(e_i,f_i)$ space — tangential to the voltage circle
— always has zero barrier resistance. The off-diagonal entry
$4\mu e_if_i/s^2$ couples the $e_i$ and $f_i$ barrier forces
in a way that depends on the operating angle, peaking at
$\delta=45^\circ$, as shown in
Fig.~\ref{fig:barrier_analysis}(b). Panel~(c) shows the coupling
ratio $|\tan(\delta)|=|bH_{ef}|/bH_{ee}$, which equals~1 at
$\delta=45^\circ$, 1.73 at $60^\circ$, and grows without bound
toward $90^\circ$ — a regime encountered in heavily loaded
networks. As $V_2\to\underline{V}_{\min}$, both formulations
produce a comparable $1/s^2$ explosion in barrier Hessian
magnitude, but the Rectangular off-diagonal coupling persists
throughout the approach to the bound, splitting the barrier force
between $e_2$ and $f_2$ rather than acting in a single coordinate
direction as in APF OPF. APF barrier coupling is identically zero
at all operating conditions.

\begin{figure}[!t]
\centering
\captionsetup{font={footnotesize}}
\includegraphics[width=\columnwidth]{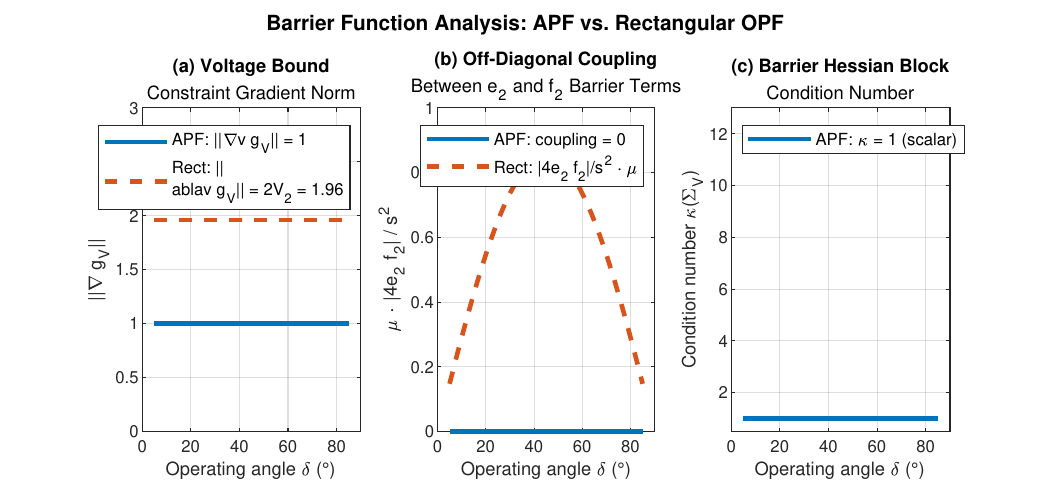}
\caption{{\color{black}Voltage bound barrier structure vs.\ operating angle:
gradient norm}}
\label{fig:barrier_analysis}
\end{figure}

\subsubsection*{Hessian and Computational Comparison}

The Cartesian formulation has one genuine advantage: a constant
power-balance Hessian. APF OPF does not achieve a constant Hessian,
but replaces the oscillatory trigonometric Hessian with smooth
rational variation. This must be weighed against the higher-degree
constraints, non-convex geometry, and
angle-dependent barrier coupling described above. To assess the net
effect, both formulations were solved using CasADi and IPOPT on the
12 benchmark systems of Table~\ref{tab:a_sensitivity} with $a=0.5$
and identical settings. APF OPF was faster in 10 of the 12 cases,
with speedups between $1.25\times$ and $3.11\times$. The largest
difference was \texttt{case588\_sdet}, where Cartesian required
4{,}498 iterations and 172~s versus 3{,}624 iterations and 107~s
for APF OPF. Objective gaps were negligible in all cases, with a
maximum of $0.0018\%$.}

\section{Conclusion}\label{sec:conclusion}

This paper presented an approximation of the OPF problem, replacing the classical trigonometric power flow kernel with an all-pass mapping equipped with a pre-rotation of angle differences. The proposed approach is effective in that it preserves the essential physical properties of the exponential kernel—unit magnitude, correct loss behavior, and the symmetry of sine and cosine—while providing smoother, non-oscillatory Jacobians and Hessians that better suit interior-point solvers. Through comprehensive implementation in both YALMIP+IPOPT and CasADi+IPOPT and extensive testing on IEEE, MATPOWER, and PGLib networks up to 10{,}480 buses, we demonstrated that the APF OPF maintains negligible optimality gaps, reproduces voltage and angle profiles and congestion patterns of the classical model, and reduces IPOPT iterations and computation time, achieving speedups. These results confirm that the proposed APF OPF formulation offers a physics-preserving and solver-friendly alternative to AC OPF.

\bibliographystyle{IEEEtran}
\bibliography{example}

@article{buason2024adaptive,
  title={Adaptive power flow approximations with second-order sensitivity insights},
  author={Buason, Paprapee and Misra, Sidhant and Watson, Jean-Paul and Molzahn, Daniel K},
  journal={IEEE Transactions on Power Systems},
  year={2024},
  publisher={IEEE}
}

@inproceedings{lofberg2004yalmip,
  title={{YALMIP}: A toolbox for modeling and optimization in {MATLAB}},
  author={Lofberg, Johan},
  booktitle={2004 IEEE international conference on robotics and automation (IEEE Cat. No. 04CH37508)},
  pages={284--289},
  year={2004},
  organization={IEEE}
}

@article{andersson2019casadi,
  title={{CasADi}: a software framework for nonlinear optimization and optimal control},
  author={Andersson, Joel AE and Gillis, Joris and Horn, Greg and Rawlings, James B and Diehl, Moritz},
  journal={Mathematical Programming Computation},
  volume={11},
  number={1},
  pages={1--36},
  year={2019},
  publisher={Springer}
}

@article{wachter2006implementation,
  title={On the implementation of an interior-point filter line-search algorithm for large-scale nonlinear programming},
  author={W{\"a}chter, Andreas and Biegler, Lorenz T},
  journal={Mathematical programming},
  volume={106},
  number={1},
  pages={25--57},
  year={2006},
  publisher={Springer}
}

@article{hiskens2001exploring,
  author  = {Hiskens, I. A. and Davy, R. J.},
  title   = {Exploring the Power Flow Solution Space Boundary},
  journal = {{IEEE} Transactions on Power Systems},
  volume  = {16},
  number  = {3},
  pages   = {389--395},
  month   = aug,
  year    = {2001}
}

@ARTICLE{hasanzadeh2025admm,
  author={Hasanzadeh, Milad and Kargarian, Amin},
  journal={IEEE Transactions on Power Systems}, 
  title={{ADMM} Enhancement Techniques for Distributed Optimal Power Flow}, 
  year={2026},
  volume={41},
  number={2},
  pages={1321-1333}}

@article{hasanzadeh2024distributed,
  title={Distributed fixed-time rotating encirclement control of linear multi-agent systems with moving targets},
  author={Hasanzadeh, Milad and Baradarannia, Mahdi and Hashemzadeh, Farzad},
  journal={Journal of the Franklin Institute},
  volume={361},
  number={11},
  pages={106970},
  year={2024},
  publisher={Elsevier}
}

@INPROCEEDINGS{hasanzadeh2025admm2,
  author={Hasanzadeh, Milad and Arthur, Fredrica and Kargarian, Amin},
  booktitle={2025 IEEE Texas Power and Energy Conference (TPEC)}, 
  title={{ADMM} Performance Enhancement Techniques for {DC} Optimal Power Flow Applications}, 
  year={2025},
  volume={},
  number={},
  pages={1-6}}

@misc{hasanzadeh2026benchmarklibrarydistributedpower,
      title={A Benchmark Library for Distributed Power System Analysis and Optimization}, 
      author={Milad Hasanzadeh and Amin Kargarian},
      year={2026},
      eprint={2506.20819},
      archivePrefix={arXiv},
}

@article{frank2016introduction,
  title={An introduction to optimal power flow: Theory, formulation, and examples},
  author={Frank, Stephen and Rebennack, Steffen},
  journal={IIE transactions},
  volume={48},
  number={12},
  pages={1172--1197},
  year={2016},
  publisher={Taylor \& Francis}
}

@article{low2014convex,
  title={Convex relaxation of optimal power flow—Part I: Formulations and equivalence},
  author={Low, Steven H},
  journal={IEEE Transactions on Control of Network Systems},
  volume={1},
  number={1},
  pages={15--27},
  year={2014},
  publisher={IEEE}
}

@article{lavaei2011zero,
  title={Zero duality gap in optimal power flow problem},
  author={Lavaei, Javad and Low, Steven H},
  journal={IEEE Transactions on Power systems},
  volume={27},
  number={1},
  pages={92--107},
  year={2011},
  publisher={IEEE}
}

@article{molzahn2019survey,
  title={A survey of relaxations and approximations of the power flow equations},
  author={Molzahn, Daniel K and Hiskens, Ian A and others},
  journal={Foundations and Trends{\textregistered} in Electric Energy Systems},
  volume={4},
  number={1-2},
  pages={1--221},
  year={2019},
  publisher={Now Publishers, Inc.}
}

@article{bienstock2019strong,
  title={Strong {NP}-hardness of {AC} power flows feasibility},
  author={Bienstock, Daniel and Verma, Abhinav},
  journal={Operations Research Letters},
  volume={47},
  number={6},
  pages={494--501},
  year={2019},
  publisher={Elsevier}
}

@article{madani2014convex,
  title={Convex relaxation for optimal power flow problem: Mesh networks},
  author={Madani, Ramtin and Sojoudi, Somayeh and Lavaei, Javad},
  journal={IEEE Transactions on Power Systems},
  volume={30},
  number={1},
  pages={199--211},
  year={2014},
  publisher={IEEE}
}

@article{bose2014equivalent,
  title={Equivalent relaxations of optimal power flow},
  author={Bose, Subhonmesh and Low, Steven H and Teeraratkul, Thanchanok and Hassibi, Babak},
  journal={IEEE Transactions on Automatic Control},
  volume={60},
  number={3},
  pages={729--742},
  year={2014},
  publisher={IEEE}
}

@inproceedings{coffrin2015strengthening,
  title={Strengthening convex relaxations with bound tightening for power network optimization},
  author={Coffrin, Carleton and Hijazi, Hassan L and Van Hentenryck, Pascal},
  booktitle={International conference on principles and practice of constraint programming},
  pages={39--57},
  year={2015},
  organization={Springer}
}

@article{jabr2006radial,
  title={Radial distribution load flow using conic programming},
  author={Jabr, Rabih A},
  journal={IEEE transactions on power systems},
  volume={21},
  number={3},
  pages={1458--1459},
  year={2006},
  publisher={IEEE}
}

@article{goodwin2025power,
  title={Power Flow Geometry and Approximation},
  author={Goodwin, Ariel and Maack, Jonathan and Sigler, Devon},
  journal={IEEE Transactions on Power Systems},
  year={2025},
  publisher={IEEE}
}

@article{stott2009dc,
  title={{DC} power flow revisited},
  author={Stott, Brian and Jardim, Jorge and Alsa{\c{c}}, Ongun},
  journal={IEEE Transactions on Power Systems},
  volume={24},
  number={3},
  pages={1290--1300},
  year={2009},
  publisher={IEEE}
}

@article{farivar2013branch,
  title={Branch flow model: Relaxations and convexification—Part I},
  author={Farivar, Masoud and Low, Steven H},
  journal={IEEE Transactions on Power Systems},
  volume={28},
  number={3},
  pages={2554--2564},
  year={2013},
  publisher={IEEE}
}

@book{nocedal2006numerical,
  title={Numerical optimization},
  author={Nocedal, Jorge and Wright, Stephen J},
  year={2006},
  publisher={Springer}
}

@article{zimmerman2010matpower,
  title={{MATPOWER}: Steady-state operations, planning, and analysis tools for power systems research and education},
  author={Zimmerman, Ray Daniel and Murillo-S{\'a}nchez, Carlos Edmundo and Thomas, Robert John},
  journal={IEEE Transactions on power systems},
  volume={26},
  number={1},
  pages={12--19},
  year={2010},
  publisher={IEEE}
}

@article{zimmerman2016matpower,
  title={Matpower 6.0 user’s manual},
  author={Zimmerman, Ray D and Murillo-S{\'a}nchez, Carlos E},
  journal={Power Systems Engineering Research Center},
  volume={9},
  pages={65--66},
  year={2016}
}

@article{bertsekas1997nonlinear,
  title={Nonlinear programming},
  author={Bertsekas, Dimitri P},
  journal={Journal of the Operational Research Society},
  volume={48},
  number={3},
  pages={334--334},
  year={1997},
  publisher={Taylor \& Francis}
}

@book{oppenheim1997signals,
  title={Signals \& Systems},
  author={Oppenheim, Alan V and Willsky, Alan S and Nawab, Syed Hamid},
  year={1997},
  publisher={Pearson Educaci{\'o}n}
}

\begin{comment}
\begin{IEEEbiography}[{\includegraphics[width=1in,height=1.25in, clip,keepaspectratio]{Milad}}]{Milad Hasanzadeh} (Graduate Student Member, IEEE) received a B.S. degree in electrical control engineering from the Sahand University of Technology, Tabriz, Iran, in 2019 and an M.S. degree in electrical control engineering in 2021 from the University of Tabriz, Tabriz, Iran. He is currently a Ph.D. student majoring in electrical and computer engineering at Louisiana State University. His research interests include multi-agent systems, finite and fixed-time design, quantum computing, and distributed optimization. Milad served as a Graduate Part-Time Instructor (GPTI) with the Department of Mechanical Engineering at Texas Tech University.
\end{IEEEbiography}
		
\begin{IEEEbiographynophoto}{Amin Kargarian} (Senior Member, IEEE) is an Associate Professor with the Department of Electrical and Computer Engineering, Louisiana State University, Baton Rouge, LA, USA. His research interests include optimization, machine learning, quantum computing, and their applications to power systems.
\end{IEEEbiographynophoto}

\end{comment}

\end{document}